\documentclass{svjour3}                     
\smartqed  
\usepackage{graphicx}

\usepackage{psfrag,epsfig,graphics}
\usepackage{amssymb}
\usepackage{amsmath}
 \usepackage{mathptmx}      
\usepackage{style_swFR}

\def\bei{\begin{itemize}}
\def\ei{\end{itemize}}

\def\citeW#1{[\citen{#1}]}

\def\vtp{\vspace{.05cm}}

\def\structure#1{#1}
\def\stmath#1{#1}
\def\aut#1{#1}
\def\alert#1{#1}
\begin{document}

\title{A short review of the theory of hard exclusive processes 
\thanks{Presented at the workshop "30 years of strong interactions", Spa, Belgium, 6-8 April 2011.}}
\subtitle{
}


\author{Samuel Wallon        
}


\institute{S. Wallon \at
              Universit\'e Pierre et Marie Curie
\&
Laboratoire de Physique Th\'eorique
 CNRS / Universit\'e Paris Sud 
Orsay \\
              \email{wallon@th.u-psud.fr}           
}

\date{Received: date / Accepted: date}

\maketitle

\begin{abstract}
We first present an introduction to the theory of hard exclusive processes. We then illustrate 
this theory by a few selected examples. The last part is devoted to the most recent developments
in the asymptotical energy limit.
\end{abstract}

\section{Introduction}
\label{Sec:intro}

\subsection{Hard processes in QCD}
\label{SubSec:hard}

Quantum chromodynamics (QCD) is the  theory of strong interaction, one of the four elementary interactions of the universe. It is a relativistic quantum field theory of  Yang-Mills 
type, with the $SU(3)$ gauge group. The  quarks and gluons elementary fields are confined inside hadrons. Nevertheless, they can be expressed as superpositions of Fock states:
\begin{itemize}

\item \structure{mesons} ($\pi, \,  \eta, f_0 ,\, \rho, \,\omega \cdots$): ${\stmath | q \bar{q} \rangle} + | q \bar{q} g \rangle+ | q q q \bar{q}\rangle + \cdots$

\item \structure{baryons} ($p, \, n, \, N, \, \Delta \cdots$): ${\stmath | q q q \rangle} + | q q q g \rangle+ | q q q q \bar{q} \rangle + \cdots$
\end{itemize}
In contrast with electrodynamics,
strong interaction increases with distance, or equivalently  {decreases when energy increases}. This phenomenum, called \alert{asymptotical freedom}, means that
the coupling satisfies ${\stmath \alpha_s(Q) \ll 1}$ for $Q \gg \Lambda_{QCD} \simeq 200$ MeV.
The natural question which then arises is how to describe and understand the internal structure of hadrons, starting from their elementary constituents, despite the confinement.
In the non-perturbative domain, the two available tools are:

\bei
\item Chiral perturbation theory: systematic expansion based on the fact that $u$ and $d$  quarks have a very small mass, the $\pi$ mass being an expansion parameter outside the chiral limit (in which these mass would be set to zero).
\vtp
\item  Discretization of  QCD on a 4-d lattice, leading to numerical simulations.
\ei

Other analytical tools have been proposed recently, among which is the AdS/QCD correspondence, a phenomenological extension of the AdS/CFT correspondence.

\begin{figure*}
\psfrag{k}{\hspace{-.3cm}$e^-$}
\psfrag{kp}{\hspace{-.1cm}$e^-$}
\psfrag{g}{\raisebox{-.2cm}{$\hspace{0cm}\gamma^*$}}
\psfrag{P}{$p$}
\psfrag{Q}{\raisebox{-.2cm}{$p$}}
\psfrag{q}{$q$}
\psfrag{p}{}
\psfrag{ppq}{}
\psfrag{pro}{\scalebox{.8}{\hspace{-.5cm}\structure{hard partonic process}}}
\centerline{\includegraphics[height=3.2cm]{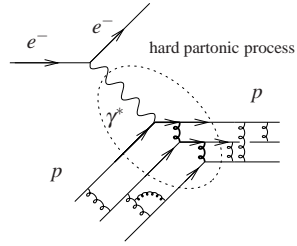} }
\caption{The partonic description of the electromagnetic proton form-factor. The partonic process at lowest order, calculable based on pertubation theory, is surrounded by a dashed line.}
\label{Fig:P_form_factor}
\end{figure*}

 Besides these tools, one may wonder whether it is possible to extract 
informations reducing
the process to interactions involving a small number of {\it partons}
(quarks, gluons), despite confinement.
This is possible if the considered process is driven by short distance phenomena, with typical distances between the interacting partons much less than $1 \, {\rm fm}$, i.e. for
$\alpha_s \ll 1.$ This is the underlying principle of 
perturbative methods.
In practice, this may be of practical use when hitting strongly enough a hadron.
This can be illustrated by the elastic scattering of an electron on a proton, introducing the proton form factor, as shown in Fig.\ref{Fig:P_form_factor}.
This description is based on following hierarchy of time scales
$$\begin{array}{l}\tau_{\mbox{\, \small electromagnetic interaction}} \sim
\tau_{\mbox{\, \small parton life time after interaction}}
 \quad \ll \tau_{\mbox{\, \small strong interaction}}
\end{array}$$
which is valid when both the virtuality $Q^2$ of the exchanged virtual photon ($\gamma^*$) and the square of the center-of-mass energy
of the $\gamma^* p$ pair are large with respect to $\Lambda_{QCD}^2\,.$

More generally, perturbative methods can be applied to any
 process governed by a hard scale, called generically {\em hard processes}. This can be the
virtuality of the electromagnetic probe: 
\bei
\item
in \structure{elastic} scattering
 $e^\pm \,p \to e^\pm \,p$
\item
 in \structure{deep inelastic scattering (DIS)} $e^\pm \,p \to e^\pm \,X$ 
\item
in \structure{deep virtual Compton scattering  (DVCS)}
$e^\pm \,p \to e^\pm \,p\, \gamma$\,.
\ei
This also
applies to $e^+ e^- \to X$ \structure{annihilation} where the hard scale is provided by the
total center of mass energy.   
In  meson photoproduction $\gamma \,p \to M\, p$ 
a large $t$-channel momentum exchange can justify the application of perturbation theory. Finally,
the hard scale can be given by the mass of a heavy bound state,   e.g. $\gamma \,p \to J/\Psi\, p$. 

\begin{figure}
\psfrag{S}{}
\begin{tabular}{ccc}
\psfrag{k}{\hspace{-.3cm}$e^-$}
\psfrag{kp}{\hspace{-.1cm}$e^-$}
\psfrag{g}{\raisebox{-.2cm}{$\hspace{0cm}\gamma^*$}}
\psfrag{P}{$p$}
\psfrag{Q}{\raisebox{-.2cm}{$p$}}
\psfrag{q}{$q$}
\psfrag{p}{}
\psfrag{ppq}{}
\psfrag{pro}{\scalebox{.7}{\hspace{-.5cm}\structure{hard partonic process}}}
\hspace{-.1cm}
\includegraphics[height=3.1cm]{partonsFormFactor.eps} 
& 
\psfrag{k}{\hspace{-.3cm}$e^-$}
\psfrag{kp}{\hspace{-.1cm}$e^-$}
\psfrag{g}{\raisebox{-.2cm}{$\hspace{0.1cm}\gamma^*$}}
\psfrag{pro}{}
\psfrag{q}{}
\psfrag{ppq}{}
\psfrag{P}{$p$}
\psfrag{X}{$X$}
\psfrag{p}{${\stmath x_{Bj}} \,p$}
\hspace{-.1cm}
\includegraphics[height=3.1cm]{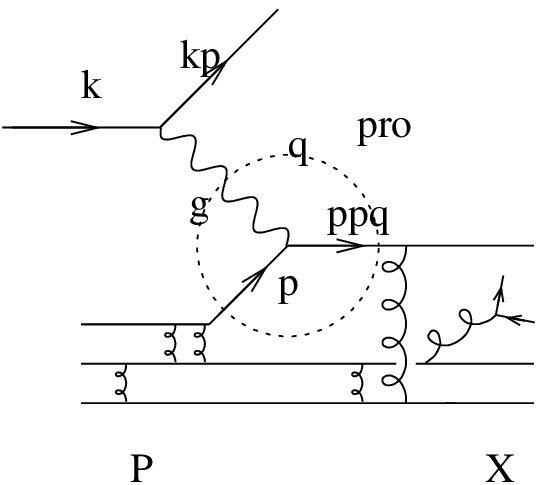}
 & 
\psfrag{k}{\hspace{-.3cm}$e^-$}
\psfrag{kp}{\hspace{-.1cm}$e^-$}
\psfrag{p}{}
\psfrag{q}{}
\psfrag{gs}{\raisebox{-.2cm}{$\hspace{0.1cm}\gamma^*$}}
\psfrag{g}{\raisebox{-.2cm}{$\hspace{0cm}\gamma$}}
\psfrag{ppq}{}
\psfrag{pro}{}
\psfrag{P}{$p$}
\psfrag{Pp}{$p$}
\hspace{-.1cm}
\includegraphics[height=3.1cm]{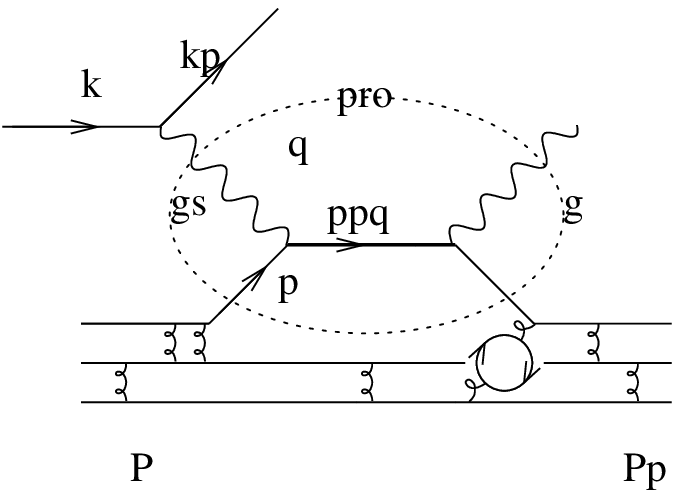}
\end{tabular}
\caption{Factorization of various hard processes. Left:
Elastic $e^- p \to e^- p$. Center: Deep inelastic scattering $e^- p \to e^- X$. Right: Deep virtual Compton scattering $e^- p \to e^- p \, \gamma$. In each case, the partonic process at lowest order is surrounded by a dashed line.}
\label{Fig:factorizations}
\end{figure}

 A precise treatment relies on \alert{factorization theorems}.
The scattering amplitude is described by the \structure{convolution} of the partonic amplitude with the non-perturbative hadronic content, as illustrated in Fig.\ref{Fig:factorizations}.

\subsection{From inclusive to exclusive processes}
\label{SubSec:inclusive_to_exclusive}

\begin{figure}
\psfrag{L}{\hspace{-.5cm}$\ln \Lambda_{QCD}^2$}
\psfrag{LQ}{$\ln Q^2$}
\psfrag{Y}{\hspace{-1cm}$Y=\ln \frac{1}{x_{Bj}}$}
\psfrag{QS}{$\ln Q_s^2(Y)$}
\psfrag{BK}{BK-JIMWLK}
\psfrag{BFKL}{BFKL}
\psfrag{DGLAP}{\hspace{-.3cm}DGLAP}
\psfrag{sat}{\raisebox{.3cm}{Saturation region}}
\psfrag{LI}{Linear regime}
\psfrag{CO}{\rotatebox{90}{Confinement}}
\centerline{\includegraphics[height=6cm]{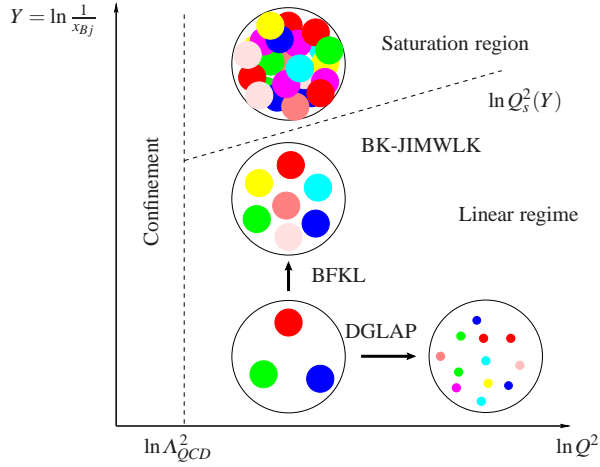}}
\caption{The various regimes governing the partonic content of the proton  in the $(\ln Q^2 \,, Y)$ plane.}
\label{Fig:DIS_phase_diagram}
\end{figure}

Historically, the  partonic proton content was first studied in DIS. In this inclusive process, the measurement of the two external kinematical variables $Q^2$ and $s_{\gamma^* p}$ give a direct access to the kinematics of the partonic process, through
\beq
\label{DIS_kinematics }
\alert{s_{\gamma^* p}} = (q_\gamma^*+ p_p)^2 = 4 \, E_{\rm c.m.}^2 \,,\quad
{\stmath Q^2} \equiv -q_{\gamma^*}^2 > 0\,, \quad
{\stmath x_{Bj}} = \frac{Q^2}{2 \, p_p \cdot q_\gamma^*} \simeq \frac{Q^2}{\alert{s_{\gamma^* p}+Q^2}}
\eq
as illustrated in Fig.\ref{Fig:factorizations} (Center).
Indeed, the two parameters ${\stmath x_{Bj}}$ and $Q^2$ have a direct interpretation in the Feynman-Bjorken mecchanism:
${\stmath x_B}$ is the proton momentum fraction carried by the scattered quark
while
${\stmath 1/Q}\ll 1/\Lambda_{QCD}$ is the transverse resolution of the photonic probe.
There are several regimes governing the evolution of perturbative content of the proton in terms of 
${\stmath x_{Bj}}$ and $Q^2$, as illustrated in Fig.\ref{Fig:DIS_phase_diagram}.
The first domain, corresponding to the 
``usual'' regime, with $s_{\gamma^* p} \sim Q^2$, for which \structure{$x_{Bj}$ is moderate} ($x_{Bj} \gtrsim .01$), is described by an 
evolution in $\alert{Q}$ governed by the QCD \structure{renormalization group}. This is the so-called Dokshitser, Gribov, Lipatov, Altarelli, Parisi (DGLAP) equation \cite{Gribov:1972ri,Altarelli:1977zs,Dokshitzer:1977sg}, which sums up terms of type
\beqa
\sum_n (\alpha_s \, \ln Q^2)^n &+& \alpha_s \, \sum_n (\alpha_s \, \ln Q^2)^n \, + \, \cdots\,. \nonumber \\
 {\rm LLQ} \qquad && \qquad{\rm NLLQ} 
\label{DGLAP_series}
\eqa
Note that this perturbative approach is based on collinear factorization, which we shall discuss further in Secs.\ref{SubSec:Extensions} and \ref{SubSec:Factorization}.

Besides this domain, in the
perturbative {\aut Regge} limit, for which  \alert{$s_{\gamma^* p} \to \infty$} i.e.
\structure{$x_{Bj} \sim Q^2/s_{\gamma^* p} \to 0$}, another evolution is expected to
deal with the stacking of partons. This leads to the Balitskii, Fadin, Kuraev, Lipatov (BFKL) equation \cite{Fadin:1975cb,Kuraev:1976ge,Kuraev:1977fs,Balitsky:1978ic},
a resummation which looks symbolically like
\beqa
\sum_n (\alpha_s \, \ln 1/x_{Bj})^n &+& \alpha_s \, \sum_n (\alpha_s \, \ln 1/x_{Bj})^n \, + \, \cdots \,. \nonumber \\
 {\rm LLx}  \qquad &&  \qquad {\rm NLLx}
\label{BFKL_series} 
\eqa
This perturbative approach is based on the $k_T$ factorization, which we shall discuss further in Secs.~\ref{SubSec:kT-factorization}. 
At very small values of $x_{Bj}$, the density of partons cannot grow for ever, and some kind of saturation phenomena should tame this growth. Its simpliest version is described by the Balitski-Kovchegov (BK) equation \cite{Balitsky:1995ub,Balitsky:1998kc,Balitsky:1998ya,Balitsky:2001re,Kovchegov:1999yj,Kovchegov:1999ua}, which realizes this saturation through ``fan diagrams'' developing from the probe toward the nucleon target, these diagrams being made of $\pom$omeron exchanges recombining through triple  $\pom$omeron vertices \cite{Bartels:1993ih,Bartels:1992ym,Bartels:1994jj,Bartels:1995kf}, a common building block of various approaches \cite{Peschanski:1997yx,Bialas:1997ig,Bialas:1997xp,Chirilli:2010mw}. 
Further extensions of these models are known under the acronym JIMWLK \cite{JalilianMarian:1997gr,JalilianMarian:1997dw,JalilianMarian:1997jx,JalilianMarian:1998cb,Kovner:2000pt,Iancu:2000hn,Iancu:2001ad,Ferreiro:2001qy,Weigert:2000gi}.

Besides these rather inclusive studies,
a very important effort is being realized 
in order to get access to the hadron structure through exclusive processes. 
Going from inclusive to exclusive processes is \alert{difficult}, since exclusive processes are rare!
This requires high luminosity accelerators and \structure{high-performance detection facilities}.
Such studies have been carried on in recent or actual experiments such as  HERA (H1, ZEUS), HERMES, JLab@6 GeV (Hall A, CLAS), BaBar, Belle, BEPC-II (BES-III). In the near future, several experiments, either already built or planned, will offer various possibilities for precise experimental studies:
LHC, COMPASS-II, JLab@12 GeV, Super-B, EIC, ILC.
Let us briefly summarize, in a non exhaustive way, the various studies which may be carried in these experimental facilities:
\bei 

	\item Proton form factor at  JLab@6 GeV and
 in the future at  PANDA (timelike proton form factor through  $p \bar{p} \to e^+ e^-$)

	\item $\gamma^* \gamma$ single-tagged channel at $e^+ e^-$ colliders  ({ \aut BaBar, Belle, BES,...}): Transition form factor $\gamma^* \gamma\to \pi,$ generalized distribution amplitudes (GDAs) in $\gamma^* \gamma\to \pi \pi,$  exotic hybrid meson production

	\item DVCS and generalized parton distributions (GPDs)
at { \aut HERA (H1, ZEUS), HERMES, JLab@6 GeV}
and in the 
future at { \aut JLab@12 GeV, COMPASS-II, EIC}, and time-like Compton scattering at  JLab@12 GeV and in ultraperipheral collisions at RHIC and LHC
\vtp

	\item Non exotic and exotic hybrid meson electroproduction: GPDs and distribution amplitudes (DAs), etc...
at
 NMC (CERN), E665 (Fermilab),  HERA (H1, ZEUS), COMPASS, HERMES, JLab

	\item Transition distribution amplitudes (TDA) { \aut(PANDA} at {\aut GSI)}

	\item Transverse momentum distributions (TMDs) ({ \aut BaBar, Belle, COMPASS, ...}) 

	\item Diffractive processes, including ultraperipheral collisions
at
 { \aut LHC} (with or without fix target), { \aut ILC}
\ei
Very important theoretical developments have been carried during the last decade. The key words are DAs, GPDs, GDAs, TDAs ... TMDs, to be explained further on. Two fundamental tool will be presented. The first one, devotted to medium energy experiments, therefore applicable at JLab, HERMES, COMPASS, BaBar, Belle, PANDA, Super-B, is the 
{\em collinear factorization}. The second one, which is specific to asymptotical energies, applies to high-energy collider experiments, like {\aut HERA, Tevatron, LHC, ILC} ({\aut EIC} and {\aut COMPASS} at the boundary), and is called {\em $k_T$-factorization}.

\subsection{Extensions from DIS}
\label{SubSec:Extensions}

\begin{figure}
\centerline{\begin{tabular}{cc}
\psfrag{ph1}{$\gamma^*$}
\psfrag{ph2}{$\gamma^*$}
\psfrag{s}{$s$}
\psfrag{t}{$t$}
\psfrag{hi}{$p$}
\psfrag{hf}{$p$}
\psfrag{Q1}{$Q^2$}
\psfrag{Q2}{$Q^2$}
\psfrag{x1}{$x$}
\psfrag{x2}{$x$}
\psfrag{GPD}{\raisebox{-.05cm}{\hspace{-.1cm}PDF}}
\psfrag{CF}{\raisebox{-.05cm}{\hspace{-.1cm}CF}}
\hspace{0cm}\raisebox{-.44 \totalheight}{\includegraphics[height=4cm]{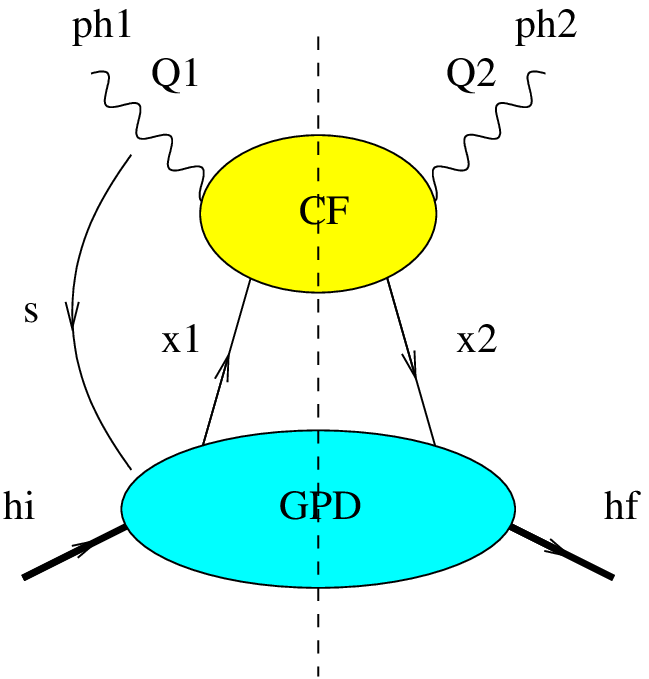}}
&
\psfrag{ph1}{$\gamma^*$}
\psfrag{ph2}{$\gamma$}
\psfrag{s}{$s$}
\psfrag{t}{$t$}
\psfrag{hi}{$p$}
\psfrag{hf}{$p'$}
\psfrag{Q1}{$Q^2$}
\psfrag{Q2}{}
\psfrag{GPD}{\raisebox{-.05cm}{\hspace{-.1cm}GPD}}
\psfrag{CF}{\raisebox{-.05cm}{\hspace{-.1cm}CF}}
\psfrag{x1}{\hspace{-.4cm}$x+\xi$}
\psfrag{x2}{$x-\xi$}
\hspace{2.5cm}\raisebox{-.44 \totalheight}{\includegraphics[height=4cm]{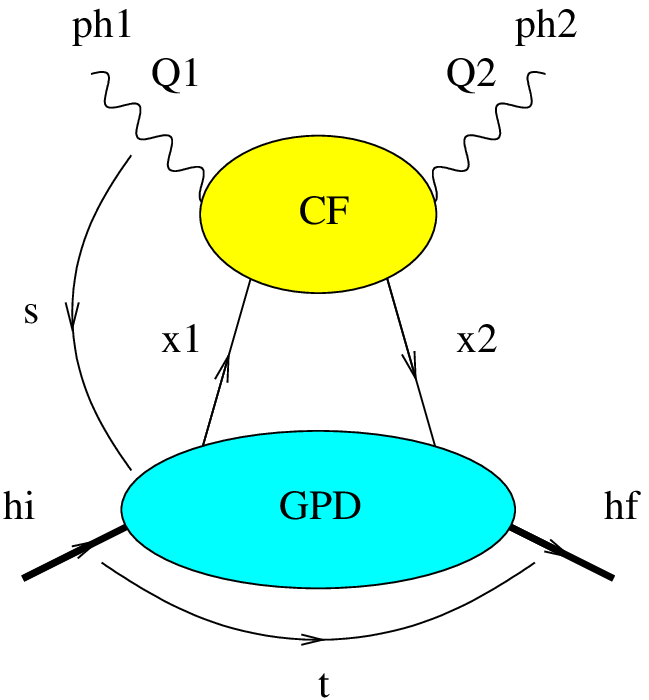}} \end{tabular}}
\caption{Factorization of DIS (left) and DVCS (right) amplitudes.}
\label{Fig:fact-DIS-DVCS}
\end{figure}

Factorizing the leptonic tensor, DIS $e^\pm p \to e^\pm X$ deals with the inclusive subprocess $\gamma^* p \to X.$
  Through optical theorem,  the total cross-section of this subprocess is related to the imaginary part of the forward Compton amplitude  $\gamma^* p \to \gamma^* p\,.$  
This amplitude can be expanded on the basis of transverse and longitudinal polarization tensors, defining the transverse and longitudinal structure functions. 
In the limit of a hard virtual photon,  this later amplitude factorizes into a hard part and a soft part, as illustrated in the left panel of Fig.~\ref{Fig:fact-DIS-DVCS}. This is a mathematical convolution (for the longitudinal momemtum fraction $x$) between coefficient functions (CFs) and parton distribution functions (PDFs), symbolically written as
\beq
\label{fact_DIS}
{\rm Im} {\cal M}_{\gamma^*p \to \gamma^*p} = CF \otimes PDF
\eq


We now consider the  virtual Compton scattering (VCS)  process 
\beq
\label{defVCS}
\gamma^*(q) \, p(p) \to \gamma^*(q') \, p(p')\,,
\eq
 which opened the way
to the introduction of non-forward parton distributions, now called GPDs\footnote{
For early reviews on GPDs, see  Refs.~\cite{Guichon:1998xv,Goeke:2001tz}).
See Refs.~\cite{Diehl:2003ny,Belitsky:2005qn} for more recent reviews.
Up-to-date reviews on models and data can be found in Refs.~\cite{Boffi:2007yc,Burkert:2007zz,Guidal:2008zza}.}.
This is a subprocess of the exclusive process
\beq
\label{eN_eNgamma}
e^\pm N \to e^\pm N \gamma \,.
\eq 
The skewness $\xi$, which caracterizes the relative amount of longitudinal momentum transfered to the nucleon, is defined in a covariant manner by
\beq
\label{def_xi}
\xi = - \frac{(q-q') \cdot(q+q')}{(p+p') \cdot(q+q')}\,.
\eq
From Eq.~(\ref{def_xi}) one deduces, in the special case of DVCS where the produced photon is real, that 
\beq
\label{xi_xBj}
\xi = \frac{x_{Bj}}{2-x_{Bj}}\,,
\eq 
which relates the skewness to the usual $x_{Bj}$ parameter. This shows in particular that at small-$x_{Bj}\,,$ typically at HERA collider (H1, ZEUS), skewness effects are rather small, and were in particular overcome in the seminal paper
\cite{Brodsky:1994kf} on diffractive electroproduction, which was devoted to HERA kinematics.

The  amplitude of the process (\ref{eN_eNgamma}) is the sum of the DVCS contribution and of the Bethe-Heitler (BH) one (where the $\gamma$ is directly emitted by the $e^\pm$). The BH process can be computed in QED, based on the measurement of proton elastic form factors. On the other hand, the DVCS
amplitude involves GPDs, which are thus in principle accessible. The squared amplitude of the process
(\ref{eN_eNgamma}) reads
\begin{equation} \label {eqn:tau}
\left| A \right|^2 
= \left| A_{{\scriptscriptstyle BH}} \right|^2 + 
\left| A_{{\scriptscriptstyle DVCS}} \right|^2 + \underbrace{
A_{{\scriptscriptstyle DVCS}} \, A_{{\scriptscriptstyle BH}}^* 
+ A_{{\scriptscriptstyle DVCS}}^* \, A_{{\scriptscriptstyle BH}}}_I\,.
\end{equation}
In practice, one can extract GPDs directly from the process (\ref{eN_eNgamma}) when the BH amplitude is negligible, which turns out to be the case at small $x_{Bj}$, a typical situation for H1 and HERA. In the more general situation when 
$x_{Bj}$ is not small, the extraction of GPDs is made easier through the study of the interference $I$ between the DVCS and the BH amplitudes. This can be done based on two generic methods: 
either by studying beam-charge asymmetries or by using beam polarization asymmetries.

The DVCS amplitude factorizes in the kinematical region $Q^2 \gg \Lambda_{QCD}$ and  $s \gg -t$ \cite{Mueller:1998fv,Radyushkin:1997ki,Ji:1998xh,Collins:1998be}: it is a convolution between  CFs and GPDs
\beq
\label{fact_DVCS}
{\cal M}_{\gamma^*p \to \gamma p} = CF \otimes GPD\,,
\eq
as illustrated at twist-2 level in the right panel of Fig.~\ref{Fig:fact-DIS-DVCS}.
A time-like version of DVCS, with an incoming on-shell photon and an outgoing time-like photon,
 factorizes and is expected to give access to the same GPDs \cite{Pire:2008ea,Pire:2011st}.

\begin{figure}
\centerline{\begin{tabular}{cc}
\psfrag{ph1}{$\gamma^*$}
\psfrag{s}{$s$}
\psfrag{t}{$t$}
\psfrag{hi}{$h$}
\psfrag{hf}{$h'$}
\psfrag{Q1}{$Q^2$}
\psfrag{GPD}{\raisebox{-.05cm}{\hspace{-.1cm}GPD}}
\psfrag{CF}{\raisebox{-.05cm}{\hspace{-.1cm}CF}}
\psfrag{DA}{\raisebox{.15cm}{\rotatebox{-55}{DA}}}
\psfrag{x1}{\hspace{-.4cm}$x+\xi$}
\psfrag{x2}{$x-\xi$}
\psfrag{z}{$u$}
\psfrag{zb}{$-\bar{u}$}
\psfrag{rho}{$\rho, \, \pi$}
\hspace{-.5cm}
\raisebox{-.44 \totalheight}{\includegraphics[height=4cm]{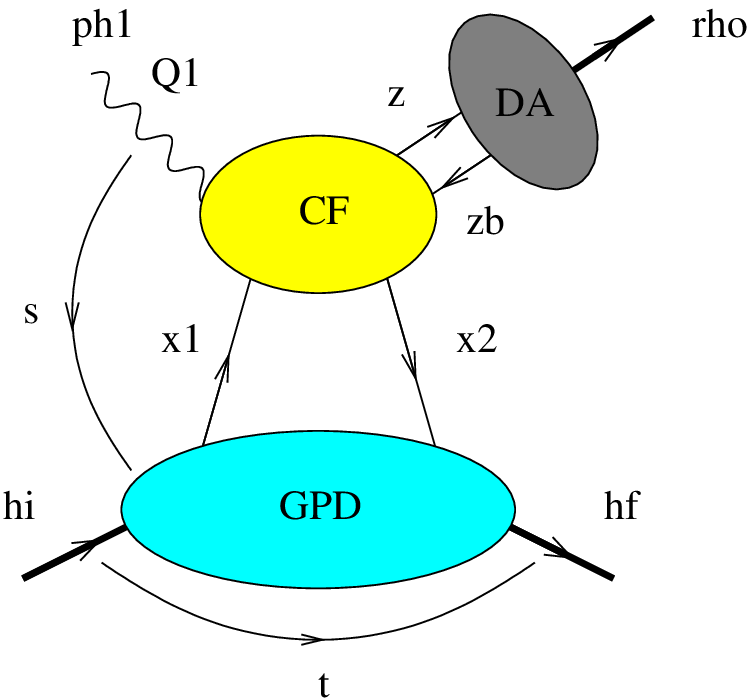}}
&
\psfrag{h1}{$h$}
\psfrag{h2}{$h'$}
\psfrag{GDA}{\raisebox{-.05cm}{\hspace{-.05cm}GDA}}
\psfrag{CF}{\raisebox{-.05cm}{\hspace{-.1cm}CF}}
\hspace{1cm}\raisebox{-.44 \totalheight}{\includegraphics[height=4cm]{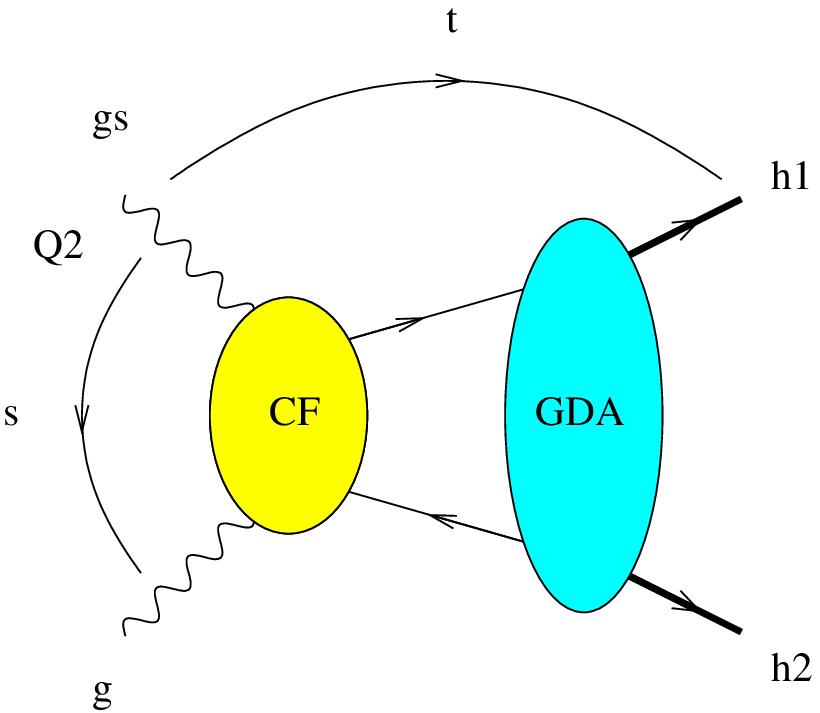}}
\end{tabular}}
\caption{Factorization of the meson electroproduction scattering amplitude (left) and of the  $\gamma^*-\gamma$ annihilation  amplitude (right).}
\label{Fig:fact-meson-GDA}
\end{figure}

Replacing the produced photon by a meson $M$, whose partonic content is described by a 
DA, meson electroproduction again factorizes  like
\cite{Collins:1996fb,Radyushkin:1997ki}
\beq
\label{fact_meson-electroproduction}
{\cal M}_{\gamma^*p \to M p} = GPD \otimes CF \otimes DA\,,
\eq
as illustrated in the left panel of Fig.~\ref{Fig:fact-meson-GDA}.
\psfrag{g}{$\gamma$}
\psfrag{gs}{$\gamma^*$}
\psfrag{s}{$s$}
\psfrag{t}{$t$}
\psfrag{Q2}{$\!\!\!Q^2$}
\psfrag{h}{\reduced hadron}
\psfrag{GDA}{\!\!\!\! \reduced  GDA}
\psfrag{CF}{\!\!\reduced  CF}
A shown in Ref.~\cite{Mueller:1998fv}
by considering the light-cone limit of
 non-local twist 2 operators, and then investigated in \cite{Diehl:1998dk,Diehl:2000uv}, the cross-process of DVCS has a factorized form in the kinematical region $Q^2 \gg \Lambda_{QCD}^2$ and  $s \ll -t$ 
\beq
\label{fact_GDA}
{\cal M}_{\gamma^* \gamma \to {\rm hadron} \, {\rm hadron}} = CF \otimes GDA\,,
\eq
where the GDAs describes the partonic content of a hadron pair.
This is illustrated in the right panel of Fig.~\ref{Fig:fact-meson-GDA}.

\begin{figure}
\centerline{\begin{tabular}{ccc}
\psfrag{ph1}{$\gamma^*$}
\psfrag{ph2}{$\gamma$}
\psfrag{s}{$s$}
\psfrag{t}{$t$}
\psfrag{hi}{$h$}
\psfrag{hf}{$h'$}
\psfrag{Q1}{$Q^2$}
\psfrag{Q2}{}
\psfrag{GPD}{\raisebox{-.05cm}{\hspace{-.1cm}GPD}}
\psfrag{CF}{\raisebox{-.05cm}{\hspace{-.1cm}CF}}
\psfrag{x1}{\hspace{-.4cm}$x+\xi$}
\psfrag{x2}{$x-\xi$}
\raisebox{-.44 \totalheight}{\includegraphics[height=4cm]{dvcs.eps}}
&
$\stackrel{t \, \to \, u}{\longrightarrow}$
&
\psfrag{hi}{$h$}
\psfrag{hf}{$h'$}
\psfrag{Q1}{$Q^2$}
\psfrag{Q2}{}
\psfrag{s}{$s$}
\psfrag{t}{$t$}
\psfrag{ph1}{$\gamma^*$}
\psfrag{ph2}{$\gamma$}
\psfrag{x1}{\hspace{-.4cm}$x+\xi$}
\psfrag{x2}{$x-\xi$}
\psfrag{GPD}{\raisebox{-.05cm}{\hspace{-.1cm}TDA}}
\psfrag{CF}{\raisebox{-.05cm}{\hspace{-.1cm}CF}}
\raisebox{-.44 \totalheight}{\includegraphics[height=4cm]{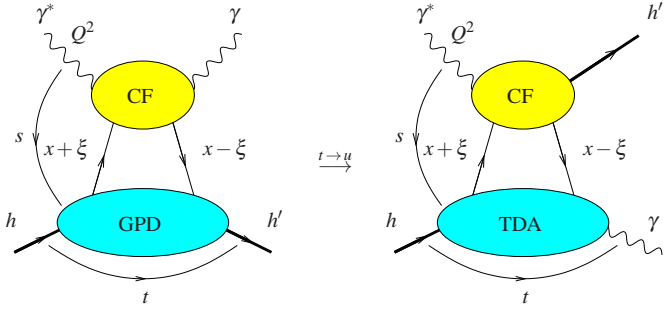}}
\end{tabular}}
\caption{
Factorized amplitude for DVCS (left) and for the crossed process backward 
DVCS in the conventional kinematics
}
\label{Fig:fact-TDA}
\end{figure}

\begin{figure}
\centerline{\scalebox{1}{\raisebox{-.44 \totalheight}{\includegraphics[height=4cm]{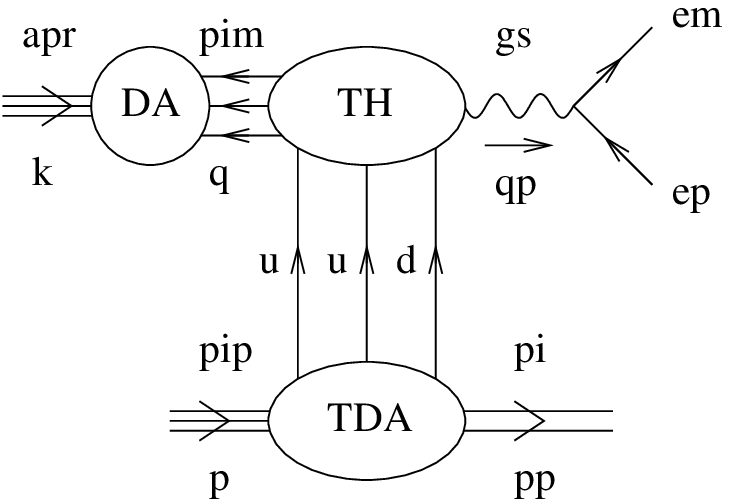}}}}
\caption{Factorization of the forward $\bar{p} p \to \gamma^* \pi $ process.}
\label{Fig:fact-TDA-example}
\end{figure}

DVCS is an extension from DIS by allowing the kinematics to be non-diagonal.
Starting from usual DVCS, further extensions are obtained by allowing the initial and final hadrons to differ. When being in the same octuplet, this leads to introduce transition GPDs. An even less diagonal quantity is naturally introduced when the baryonic numbers of the initial and final hadron differ, by $t \leftrightarrow u$ crossing from DVCS, leading to the introduction \cite{Pire:2004ie,Pire:2005ax} of the TDA of the hadron to a photon, as shown in Fig.~\ref{Fig:fact-TDA}. This can be further extended by replacing the outgoing $\gamma$ by any hadronic state. 
\psfrag{p}{$p$}  
\psfrag{pp}{$p'$}
\psfrag{q}{$q$}
\psfrag{qp}{$q'$}
\psfrag{pip}{$p$}
\psfrag{pim}{$\pi^-$}
\psfrag{pr}{$p$}
\psfrag{apr}{$\bar p$}
\psfrag{g}{$\gamma$}
\psfrag{gs}{$\gamma^*$}
\psfrag{u}{$u$}
\psfrag{db}{$\bar d$}
\psfrag{ep}{$e^+$}
\psfrag{em}{$e^-$}
\psfrag{d}{$d$} 
\psfrag{a}{$a$}
\psfrag{b}{$b$}
\psfrag{pi}{$\pi$}
\psfrag{k}{$k$}
\psfrag{DA}{$\,\,DA$}
\psfrag{TDA}{$TDA$}
\psfrag{TH}{$T_H$}
\psfrag{q}{}
\psfrag{pim}{}
As an example, the $p \to \pi$ TDA could be studied at PANDA, in the forward $\bar{p} p \to \gamma^* \pi $ scattering \cite{Lansberg:2007se}, as shown in Fig.~\ref{Fig:fact-TDA-example}. 

\subsection{Factorization}
\label{SubSec:Factorization}

\begin{figure}
\begin{center}
\scalebox{.95}{\begin{tabular}{ccc}
\hspace{.3cm}\raisebox{-.44 \totalheight}{\includegraphics[height=4.5cm]{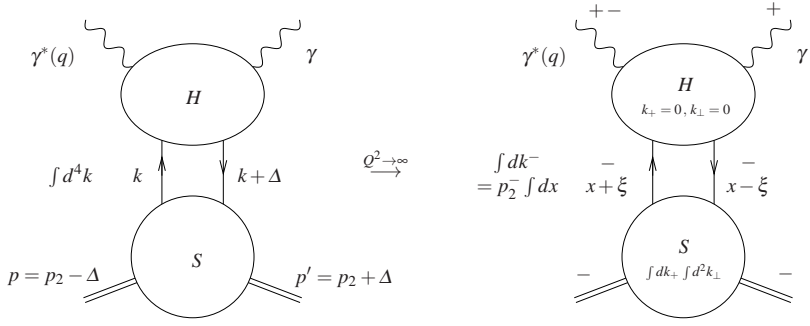}} 
& \hspace{-.1cm} $\stackrel{Q^2 \to \infty}{\longrightarrow}$ \hspace{.5cm}&
\psfrag{H}[cc][cc]{\scalebox{1}{\begin{tabular}{c}$\alert{H}$\\ \scalebox{.7}{$k_+=0\, , \,k_\perp=0$} \end{tabular}}}
\psfrag{S}[cc][cc]{\scalebox{1}{\begin{tabular}{c}${\stmath S}$\\ \scalebox{.7}{$\int d k_+\,  \int d^2 k_\perp$} \end{tabular}}}
%
\psfrag{pl}[rc][Br]{$ \begin{array}{c} \int dk^{\alert{-}}\\  =p_2^{\alert{-}} \int dx \end{array} \quad \begin{array}{c}\alert{-}\\  x+\xi \end{array}$}
\psfrag{pr}[cc][Bc]{$\begin{array}{c}\alert{-}\\  x-\xi \end{array}$}
\psfrag{pp}[rc][rc]{$\alert{-}$}
\psfrag{ppp}[cc][lc]{$\alert{-}$}
\psfrag{pmq}[cc][cc]{$+ \, -$}
\psfrag{mu}[cc][cc]{$+$}
\hspace{1.6cm}
\raisebox{-.44 \totalheight}{\includegraphics[height=4.5cm]{factGPD_HsD.eps}} 
\end{tabular}}
\end{center}
\caption{Factorization of DVCS in momentum space at large $Q^2.$}
\label{Fig:Factorization-DVCS-momentum}
\end{figure}

Factorization relies on two steps: the first one is based on momentum factorization, based on the light-cone dominance in the $Q^2 \to \infty$ limit. The natural frame to set up this factorization is the Sudakov decomposition, introducing two light-cone directions $p_1$ and $p_2$
\beq
p_1 =\frac{\sqrt{s}}{2}(1,0_\perp,1)\,, \quad p_2 =\frac{\sqrt{s}}{2}(1,0_\perp,-1)
\label{Sudakov-Basis}
\eq
with $2 \, p_1 \cdot p_2=s \sim s_{\gamma^*p}$.
Any four-vector is then  expanded according to
\beq
\label{expansion-Sudakov}
\begin{array}{ccccccc}
\alert{k} &=& \alpha \, p_1 &+& \beta \, p_2 &+& k_\perp \,.\\
& & \stmath{+} & & \alert{-} & & \stmath{\perp}
\end{array}
\eq
At large $Q^2$, considering the  momentum $k$ of the parton connecting the hard part $H$ with the soft part $S$, the hard part only depends on the component of $k$ along the incident hadron (denoted as the  $-$ component). In this approximation, the amplitude is then the convolution with respect to the $-$ fraction of the hard and soft part, as illustrated in Fig.~\ref{Fig:Factorization-DVCS-momentum}, and reads symbolicaly
\psfrag{H}[cc][cc]{\hspace{-.1cm} \vspace{.3cm} $\alert{ H}$} 
\psfrag{S}[cc][cc]{$\hspace{.1cm} {\stmath S}$} 
\psfrag{gas}[rc][lc]{$\gamma^*(q)$}
\psfrag{g}[cc][cc]{$\gamma$}
\psfrag{D}[cc][cc]{}
\psfrag{pp}[rt][rt]{$p=p_2-\Delta$}
\psfrag{ppp}[lt][lt]{$p'=p_2+\Delta$}
%
\psfrag{pl}[rc][Br]{$\int d^4k \qquad k$}
\psfrag{pr}[cc][Bc]{$\ k+\Delta$}
\psfrag{pmq}[cc][cc]{$+ \, -$}
\psfrag{mu}[cc][cc]{$+$}
\psfrag{pmq}[cc][cc]{}
\psfrag{mu}[cc][cc]{}
\beq
\int d^4 k \,\, S(k,\, k+\Delta) \, H(q,\, k,\, k+\Delta) = \int dk^- \stmath{\int dk^+ d^2 k_\perp \, S(k,\, k+\Delta)}  \alert{H(q, \,k^-,\, k^- +\Delta^-)} 
\label{SH-factorization}
\eq
The second step is to perform the factorization with respect to quantum numbers, in accordance to $C$, $P$ and $T$ parity which select the allowed structures when performing Fierz decomposition in $t$ channel (among the 16 Dirac matrices in the case of quark exchange).
\psfrag{gas}[cc][lc]{$\!\!\!\!\gamma^*(q)$}
\psfrag{M}[cc][lc]{$\,\,\,\alert{M}$}
\psfrag{P}[cc][lc]{$\,\,\,\stmath{\Psi}$}
\psfrag{pp}[cB][lc]{$p$}
\psfrag{ppp}[lB][lc]{$\,\,\,\,\, \, p'$}
\psfrag{r}[ct][ct]{$\,  \stmath{\rho}$}
\psfrag{V}[cB][cc]{$p_\rho$}
%
%
\psfrag{pmq}[cc][cc]{}
\psfrag{mu}[Bc][lc]{$\hspace{-.5cm}\int d^4 \ell \quad \ \ell$}
\psfrag{md}[Bc][cc]{$\ell-p_\rho$}
\begin{figure}[h]
\scalebox{.95}{\begin{tabular}{ccc}
\hspace{0.3cm}\raisebox{-.44 \totalheight}{\includegraphics[height=2.4cm]{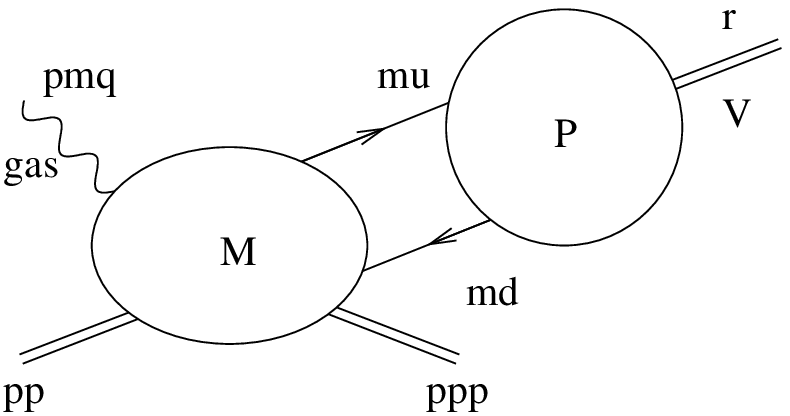}} 
& \hspace{-.2cm}  \raisebox{-.44 \totalheight}{$\stackrel{Q^2 \to \infty}{\longrightarrow}$} \hspace{.1cm} &
\psfrag{M}[cc][cc]{\scalebox{1}{\begin{tabular}{c}$\alert{M}$\\ \scalebox{.7}{$\ell_-=0\, , \,\ell_\perp=0$} \end{tabular}}}
\psfrag{P}[cc][cc]{\scalebox{1}{\begin{tabular}{c}${\stmath S}$\\ \scalebox{.7}{$\int d \ell_-\,  \int d^2 \ell_\perp$} \end{tabular}}}
\psfrag{gas}[cc][lc]{${\stmath +} \, \alert{-}$}
\psfrag{pmq}[cc][cc]{}
%
\psfrag{mu}[Bc][lc]{\hspace{-1.2cm}\raisebox{.35 \totalheight}{$ \begin{array}{cc}\footnotesize{\begin{array}{c} \int d\ell^+ \\  =p_1^+ \int du \end{array}}& \ u \, \, {\stmath +}\end{array}$}}
\psfrag{md}[Bc][cc]{$\!-\bar{u} \,\,\,{\stmath +}$}
\psfrag{r}[ct][ct]{$\,  \stmath{\rho}$}
\psfrag{V}[cB][cc]{${\stmath +}$}
%
%
\psfrag{pp}[cB][rc]{$\alert{-}$}
\psfrag{ppp}[cB][lc]{$\alert{-}$}
\raisebox{-.44 \totalheight}{\includegraphics[height=3.2cm]{factM_DA.eps}}
\end{tabular}}
\caption{Collinear factorization of $\rho$-electroproduction.}
\label{Fig:factorization-DA}
\end{figure}

The case of $\rho$-meson production involves a second collinear factorization. Indeed the  $\rho$-meson is described by its wave function 
$\stmath\Psi$ which reduces for \structure{hard processes} to its DA
 \cite{Chernyak:1977as,Chernyak:1977fk,Chernyak:1980dk,Chernyak:1980dj,Lepage:1980fj}, originally introduced in the case of form factors, as (denoting $l^+ = u \, p_2$)
\beq
\label{def-DA}
\Phi(u,\muF)=\stmath{\int d\ell^-\hspace{-.3cm} \int\limits^{\scalebox{.6}{$|\ell_\perp^2| < \muF$}}
d^2 \ell_\perp \, \Psi(\ell,\, \ell-p_\rho)}
\eq
where $\mu_F$ is the factorization scale. For large $Q^2$, factorization symbolicaly reads
\beqa
\label{factorization-DA}
\int d^4 \ell \,\,  M(q,\, \ell ,\, \ell -p_\rho) 
 \Psi(\ell,\, \ell-p_\rho) &\!=&\! \int d\ell^+ \alert{M(q; \,\ell^+,\, \ell^+ -p_\rho^+)}
\stmath{\int d\ell^- \!\!\!\!\!\int\limits^{\scalebox{.5}{$|\ell_\perp^2| < \mu_F^2$}}\!\!
d^2 \ell_\perp \, \Psi(\ell,\, \ell-p_\rho)} \nonumber \\
&\!=&\! p_\rho^+ \int du \, \alert{M(q; \,u \, p_\rho^+,\, - \bar{u} p_\rho^+;\mu_F)} \, \Phi(u,\mu_F)\,.
\eqa
as illustrated in Fig.~\ref{Fig:factorization-DA}. 
The arbitrariness of the amplitude with respect to $\mu_F$ leads to the Efremov, Radyushkin, Brodsky, Lepage (ERBL) equations for the DAs \cite{Farrar:1979aw,Lepage:1979zb,Efremov:1979qk}.
\psfrag{M}[cc][lc]{$\,\,\,\stmath{\Psi}$}
\psfrag{S}[cc][lc]{$\stmath{S}$}
\psfrag{H}[cc][lc]{$\alert{H}$}
\psfrag{gas}[cc][lc]{$\gamma^*(q)$}
\psfrag{g}[cc][cc]{$\gamma$}
\psfrag{D}[cc][cc]{}
\psfrag{pp}[cc][cc]{$p$}
\psfrag{ppp}[cc][cc]{$p'$}
\psfrag{r}[cc][cc]{}
\psfrag{V}[cc][cc]{$p_\rho$}
\psfrag{gas}[rc][lc]{$\gamma^*(q)$}
\psfrag{g}[cc][cc]{$\gamma$}
\psfrag{D}[cc][cc]{}
\psfrag{pp}[rt][rt]{\footnotesize{$p=p_2-\Delta$}}
\psfrag{ppp}[lt][lt]{\footnotesize{$p'=p_2+\Delta$}}
%
%
%
%
\psfrag{pmq}[cc][cc]{}
\psfrag{mu}[Bc][lc]{$\hspace{-.4cm}\int d^4 \ell \quad \ell$}
\psfrag{md}[Bc][cc]{$\ell-p_\rho$}
%
\psfrag{pl}[rc][Br]{$\int d^4k \qquad k$}
\psfrag{pr}[cc][Bc]{$\ k+\Delta$}
\begin{figure}[h]
\hspace{1cm}\scalebox{.99}{
\begin{tabular}{ccc}
\scalebox{.9}{\raisebox{-.44 \totalheight}{\includegraphics[height=4.5cm]{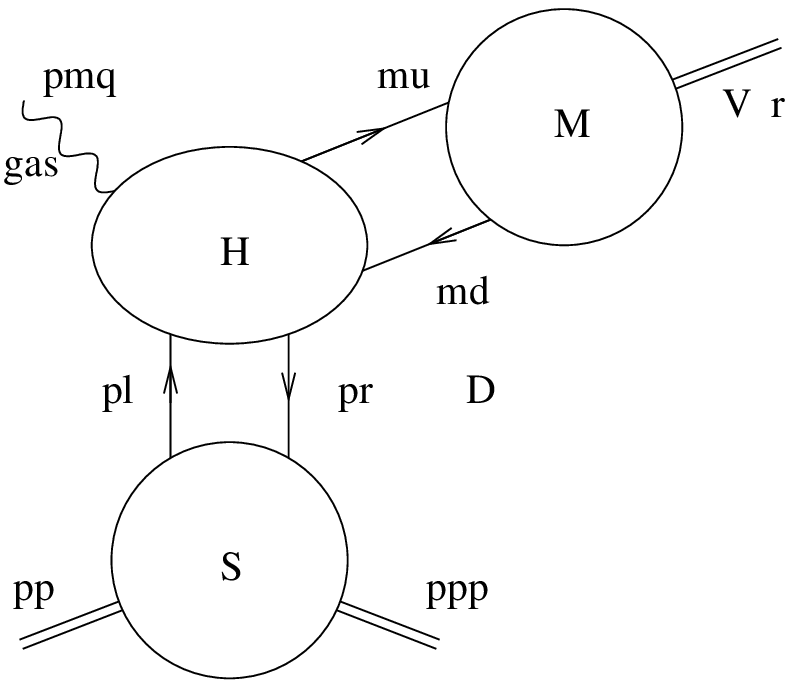}}}
& \hspace{-.8cm} \raisebox{-.44 \totalheight}{$\stackrel{Q^2 \to \infty}{\longrightarrow}$}  &
\psfrag{H}[cc][cc]{\raisebox{.2cm}
{\scalebox{1}{\begin{tabular}{c}$\alert{H}$\\ \scalebox{.48}{$\ell_-=0\, , \,\ell_\perp=0$} \\
\scalebox{.48}{$k_+=0\, , \,k_\perp=0$}
\end{tabular}}}}
\psfrag{M}[cc][cc]{\scalebox{1}{\begin{tabular}{c}${\stmath S}$\\ \scalebox{.7}{$\int d \ell_-\,  \int d^2 \ell_\perp$}\end{tabular}}}
\psfrag{S}[cc][cc]{\scalebox{1}{\begin{tabular}{c}${\stmath S}$\\ \scalebox{.7}{$\int d k_+\,  \int d^2 k_\perp$} \end{tabular}}}
\psfrag{gas}[cc][lc]{}
\psfrag{pl}[rc][Br]{$\int dx \quad \begin{array}{c}\alert{-}\\  x+\xi \end{array}$}
\psfrag{pr}[cc][Bc]{$\begin{array}{c}\alert{-}\\  x-\xi \end{array}$}
\psfrag{pp}[rc][rc]{$\alert{-}$}
\psfrag{ppp}[cc][lc]{$\alert{-}$}
%
%
\psfrag{pmq}[cc][cc]{${\stmath +} \, \alert{-}$}
\psfrag{mu}[Bc][lc]{\hspace{-1.2cm}\raisebox{.35 \totalheight}{$ \begin{array}{cc} \int du &  u \ \, \stmath{+}\end{array}$}}
\psfrag{md}[Bc][cc]{$\!-\bar{u} \ \,\stmath{+}$}
\psfrag{V}[cc][cc]{$\stmath{+}$}
\hspace{.7cm}
\scalebox{.9}{\raisebox{-.44 \totalheight}{\includegraphics[height=4.5cm]{factGPD_H_DAsD.eps}}}
\end{tabular}}
\caption{Momentum space factorization of $\rho$-electroproduction.}
\label{Fig:factorization-DA-GPD}
\end{figure}

The scattering  amplitude for meson electroproduction has the
fully factorized form, shown in the left panel of Fig.~\ref{Fig:factorization-DA-GPD},
\beqa
\label{factorization-DA-GPD}
&&\int d^4 k \, d^4 \ell\,\, \stmath{S(k,\, k+\Delta)} \alert{H(q;\, k,\, k+\Delta)} \stmath{\Psi(\ell,\, \ell-p_\rho)}= 
p^- p_\rho^+ \int dx \, du\\
&&\!\!\!\!\!\!\times \!\!\left[\stmath{\int \!\!dk^+\hspace{-.3cm} \int\limits^{\scalebox{.6}{$|k_\perp^2| < \muFb$}} \hspace{-.3cm} d^2 k_\perp \, S(k,\, k+\Delta)}\right]  \alert{ H(q; \,(x+\xi)p^- ,\, (x-\xi) p^- ; u \, p_\rho^+, \, -\bar{u} \, p_\rho^+; \mu_{F_1} ; \mu_{F_2})}  \Phi(u,\mu_{F_1})\,.\nonumber
%
\eqa
\begin{figure}[h]
\hspace{-.2cm}
\psfrag{q1}[cc][cc]{$Q^2$}
\psfrag{pl}[rc][Br]{}
\psfrag{pr}[cc][Bc]{}
\psfrag{pp}[rc][rc]{}
\psfrag{ppp}[cc][lc]{}
%
\psfrag{n}[cc][cc]{\structure{$\Gamma$}}
\psfrag{p1}[cc][cc]{\structure{$\Gamma$}}
\psfrag{n}[cc][cc]{\alert{$\Gamma'$}}
\psfrag{p}[cc][cc]{\alert{$\Gamma'$}}

\psfrag{Tda}[cc][cc]{$TDA_H$}
\psfrag{Th}[cc][cc]{$H$}
\psfrag{Da}[cc][cc]{DA}
\psfrag{S}[cc][cc]{GPD}
\centerline{\scalebox{.8}
{
$\begin{array}{c}
\begin{array}{cc}
\raisebox{-0.44 \totalheight}{\epsfig{file=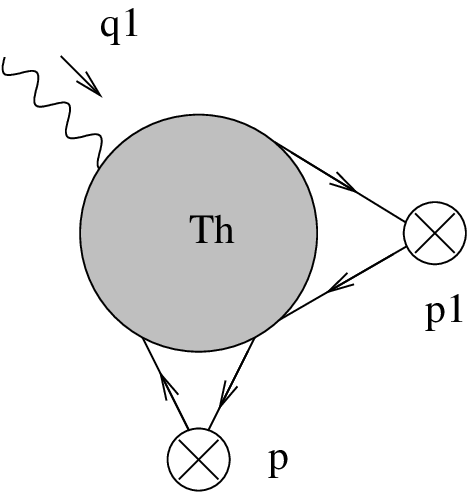,width=2.8cm}}&
\raisebox{-0.27\totalheight}
{\psfrag{r}[cc][cc]{$\qquad \quad M(p,\lambda)$}
\psfrag{pf}[cc][cc]{\raisebox{-1.5 \totalheight}{\hspace{-.1cm}$\Gamma$}}\epsfig{file=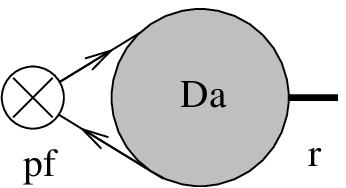,width=2cm}}
\end{array}\\
\\
\begin{array}{cc}
\hspace{-2.15cm}\raisebox{-0.44 \totalheight}{\epsfig{file=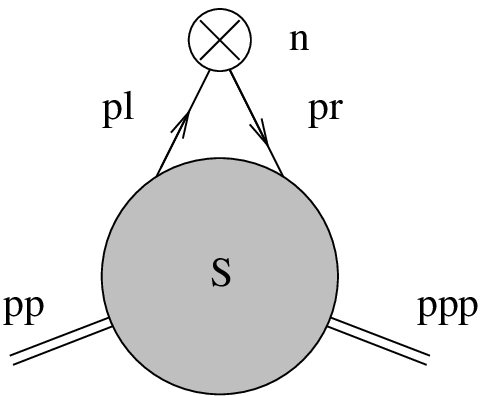,width=2.8cm}}&
\end{array}
\end{array}
$}}
\caption{Factorization of $\rho$-electroproduction including quantum numbers. Crosses symbolize $\Gamma$ matrices.}
\label{Fig:factorization-DA-GPD-blocks}
\end{figure}
After completing momentum and quantum number factorization, we have thus been led to  introduce three building blocks entering Fig.~\ref{Fig:factorization-DA-GPD-blocks}.
These are
\psfrag{q1}[cc][cc]{$\hspace{.2cm}Q^2$}
%
\psfrag{pl}[rc][Br]{}
\psfrag{pr}[cc][Bc]{}
\psfrag{pp}[cc][lc]{\raisebox{1 \totalheight}{$\hspace{-.3cm}N(p)$}}
\psfrag{ppp}[cc][lc]{\raisebox{1 \totalheight}{$\hspace{.4cm}N(p')$}}
%
\psfrag{n}[cc][cc]{\structure{$\Gamma$}}
\psfrag{p1}[cc][cc]{\structure{$\Gamma$}}
\psfrag{n}[cc][cc]{\alert{$\Gamma'$}}
\psfrag{p}[cc][cc]{\alert{$\Gamma'$}}
\psfrag{Tda}[cc][cc]{$TDA_H$}
\psfrag{Th}[cc][cc]{$H$}
\psfrag{Da}[cc][cc]{DA}
\psfrag{S}[cc][cc]{GPD}
\beqas
\hspace{.5cm}\raisebox{-0.46 \totalheight}{\epsfig{file=THLorder.eps,width=\scalingm}}
\hspace{.2cm}
&=& \hspace{.2cm}
\raisebox{-0.46 \totalheight}{\epsfig{file=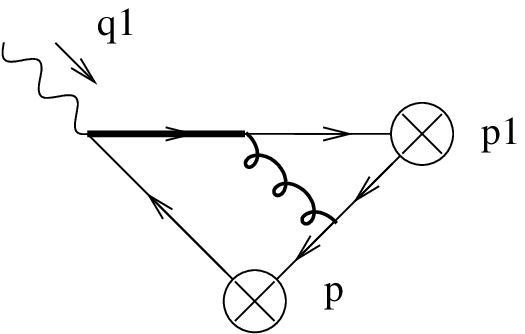,width=\scalingm}}
+
\psfrag{q1}[cc][cc]{$\hspace{.35cm}Q^2$}
\raisebox{-0.46 \totalheight}{\epsfig{file=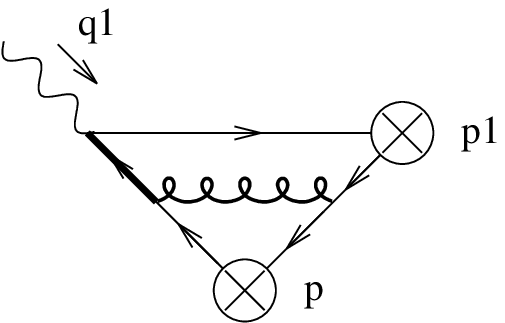,width=\scalingm}} \quad \hbox{\structure{\footnotesize hand-bag diagrams}}
\\
\\
\hspace{-.1cm}\raisebox{-0.46 \totalheight}{\psfrag{r}[cc][cc]{$\hspace{.7cm} M(p,\lambda)$}
\psfrag{pf}[cc][cc]{\raisebox{-3\totalheight}{\structure{$\Gamma$}}}
\epsfig{file=DA.eps,width=2cm}}  \hspace{0.4cm} &\hspace{0cm}=& \hspace{.4cm}\langle M(p,\lambda)  | {\cal O} (\Psi, \, \bar{\Psi}, \, A)  | 0 \rangle \quad 
\begin{array}{l} 
 \hbox{\structure{\footnotesize matrix element of a \alert{non-local light-cone}}}
\\
  \hbox{\structure{\footnotesize operator}}
\end{array}
\\
\\
\hspace{-.6cm}\raisebox{-0.46 \totalheight}{\psfrag{r}[cc][cc]{$\qquad \quad M(p,\lambda)$}
\psfrag{pf}[cc][cc]{\raisebox{-3\totalheight}{$\Gamma$}}
\epsfig{file=SoftDVCS_light_grey.eps,width=2cm}} \hspace{0.2cm} & =& \hspace{.2cm}\langle N(p')  | {\cal O}' (\Psi, \, \bar{\Psi}, \, A)  | N(p) \rangle  \ \
\begin{array}{l} 
 \hbox{\structure{\footnotesize matrix element of a \alert{non-local light-cone}}}
\\
  \hbox{\structure{\footnotesize operator}}
\end{array}
\eqas

\subsection{GPDs at twist 2}
\label{SubSec:GPDs-at-twist-2}

\begin{figure}[h]
\begin{center}
     \leavevmode
     \epsfxsize=.95\textwidth

     \epsffile{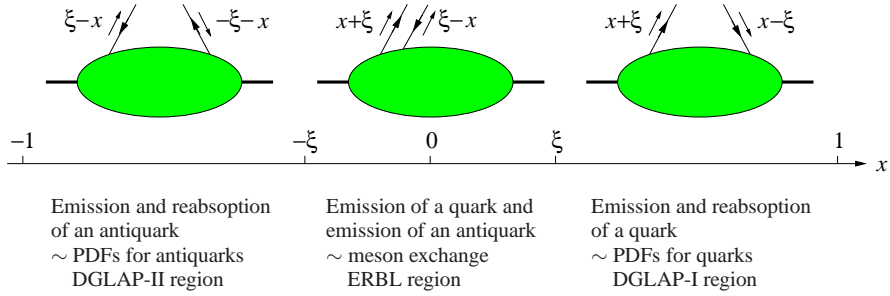}
\end{center}

\begin{tabular}{ccc}
\hspace{.4cm}
\begin{tabular}{ll}
\structure{Emission and reabsoption} \\
\structure{of an antiquark}\\
\structure{$\sim$ PDFs for antiquarks}\\
\quad {\aut DGLAP}-II region
\end{tabular}
&
\hspace{-.2cm}
\begin{tabular}{ll}
\structure{Emission of a quark and} \\
\structure{emission of an antiquark}\\
\structure{$\sim$ meson exchange}\\
\quad {\aut ERBL} region
\end{tabular}
&
\hspace{-.2cm}
\begin{tabular}{ll}
\structure{Emission and reabsoption} \\
\structure{of a quark}\\
\structure{$\sim$ PDFs for quarks}\\
\quad {\aut DGLAP}-I region
\end{tabular}
\end{tabular}
\caption{\label{fig:regions_GPD} The  partonic interpretation of GPDs (Fig. from Ref.\cite{Diehl:2003ny}).}
\label{Fig:GPDs-twist-2}
\end{figure}

The GPDs have a simple physical interpretation at twist 2, illustrated in Fig.~\ref{Fig:GPDs-twist-2}, based on density number operators \cite{Ji:1998pc,GolecBiernat:1998ja,Diehl:2000xz}.
As for DAs, the arbitrariness in the choice of $\mu_{F_2}$ leads to evolution equations for GPDs,
called ERBL-DGLAP equations \cite{Bukhvostov:1984as,Bukhvostov:1985rn,Dittes:1988xz,Mueller:1998fv}, which are extensions of the  ERBL \cite{Farrar:1979aw,Lepage:1979zb,Efremov:1979qk} and DGLAP \cite{Gribov:1972ri,Altarelli:1977zs,Dokshitzer:1977sg} evolution equations.

For quarks, one should distinguish two kinds of GPDs. The exchanges
\structure{without helicity flip} involve \structure{chiral-even} \alert{$\Gamma'$} matrices, and define 4 chiral-even GPDs: 
$H^q$ (reducing to the \structure{PDF} $q$ in the limit $\xi=0,$ $t=0$), 
 $E^q$, $\tilde{H}^q$ (which is the \structure{polarized PDF} $\Delta q$
 in the limit $\xi=0,$ $t=0$) and $\tilde{E}^q$, defined by
\beqa
  \label{def-GPD-quark-even}
F^q &=&
\frac{1}{2} \int \frac{d z^+}{2\pi}\, e^{ix P^- z^+}
  \langle p'|\, \bar{q}(-\half z)\, {\alert{\gamma^-}} q(\half z) 
  \,|p \rangle \Big|_{z^-=0,\, \tvec{z}=0}
\nonumber \\
&=& \frac{1}{2P^-} \left[
  \alert{H^q}(x,\xi,t)\, \bar{u}(p') \gamma^- u(p) +
  \alert{E^q}(x,\xi,t)\, \bar{u}(p') 
                 \frac{i \,\sigma^{-\alpha} \Delta_\alpha}{2m} u(p)
  \, \right] ,
\nonumber \\
\tilde{F}^q &=&
\frac{1}{2} \int \frac{d z^+}{2\pi}\, e^{ix P^- z^+}
  \langle p'|\, 
     \bar{q}(-\half z)\, \alert{\gamma^- \gamma_5}\, q(\half z)
  \,|p \rangle \Big|_{z^-=0,\, \tvec{z}=0}
\nonumber \\
&=& \frac{1}{2P^-} \left[
 \alert{\tilde{H}^q}(x,\xi,t)\, \bar{u}(p') \gamma^- \gamma_5 u(p) +
  \alert{\tilde{E}^q}(x,\xi,t)\, \bar{u}(p') \frac{\gamma_5 \,\Delta^-}{2m} u(p)
  \, \right] .
\eqa
The exchanges
	{with helicity flip} involve \structure{chiral-odd} \alert{$\Gamma'$}  matrices, leading to the
\alert{4 chiral-odd GPDs}
$H^q_T$ (the \structure{quark transversity PDFs} $\Delta_T q$ when $\xi=0,$ $t=0$), \alert{$E^q_T$, $\tilde{H}^q_T$, $\tilde{E}^q_T$}, defined by
\beqa
\label{def-GPD-quark-odd}
&&\hspace{-.5cm}\frac{1}{2} \int \frac{d z^+}{2\pi}\, e^{ix P^- z^+}
  \langle p'|\, 
     \bar{q}(-\half z)\, i \, \alert{\sigma^{-i}}\, q(\half z)\, 
  \,|p \rangle \Big|_{z^-=0,\, \tvec{z}=0} 
 \\
&&\hspace{-.5cm}= \frac{1}{2P^-} \bar{u}(p') \left[
 \alert{H_T^q}\, i \sigma^{-i} +
  \alert{\tilde{H}_T^q}\, \frac{P^- \Delta^i - \Delta^- P^i}{m^2} +
  \alert{E_T^q}\, \frac{\gamma^- \Delta^i - \Delta^- \gamma^i}{2m} +
  \alert{\tilde{E}_T^q}\, \frac{\gamma^- P^i - P^- \gamma^i}{m}
  \right] u(p) \, . \nonumber
\eqa
Analogously,  there are
 \alert{4 gluonic GPDs} \structure{without helicity flip:} 
\alert{$H^g$} (it is the \structure{PDF} $x \, g$ in the limit $\xi=0$, $t=0$),
\alert{$E^g$},
\alert{$\tilde{H}^g$} (it is the \structure{polarized PDF} $x \, \Delta g$ when  
$\xi=0$, $t=0$) and
\alert{$\tilde{E}^g$}; and 
\alert{4 gluonic GPDs} \structure{with helicity flip:} 
\alert{$H^g_T$},
\alert{$E^g_T$},
\alert{$\tilde{H}^g_T$} and
\alert{$\tilde{E}^g_T$}
(there is no forward limit reducing to gluons PDFs here: a change of  2 units of helicity cannot be  compensated by a spin 1/2 target).

\subsection{Selection rules and factorization status}
\label{SubSec:selection-rules_factorization_status}

The selection rule for the meson electroproduction can be obtained in a simple manner.
Since for a massless particle 
 chirality = + (resp. -) helicity for a (anti)particule 
and based on the fact that \structure{QED and QCD vertices are chiral even} (no chirality flip during the interaction), one deduces\footnote{This is the same reason which explains the vanishing of $F_L$ in DIS.} that 
the total helicity of a $q \bar{q}$ pair produced by a $\gamma^*$ should be 0.
Therefore, the helicity of the $\gamma^*$ equals the $z$ projection of the $q \bar{q}$ angular momentum $L^{q \bar{q}}_z$.
In the pure collinear limit (i.e. twist 2), the $q \bar{q}$ does not carry any angular momentum: $L^{q \bar{q}}_z=\,0\,.$ Thus
the $\gamma^*$ is longitudinally polarized.
Additionaly, 
at $t=0$ there is no source of orbital momentum from the proton coupling, which implies
that 
the helicity of the meson and of the photon should be identical.
In the collinear factorization approach, the extension to $t \neq 0$ changes nothing from the hard part side, 
the only dependence with respect to $t$ being encoded in the non-perturbative correlator which defines the GPDs. This implies that 
  the above selection rule remains true. Thus, 
only   2 transitions are possible (this is the so-called $s-$channel helicity conservation (SCHC)):
 $\gamma^*_L \to \rho_L$, for which QCD factorization \alert{ holds at t=2} at any order 
(i.e. LL, NLL, etc...) \cite{Collins:1996fb} and 
$\gamma^*_T \to \rho_T$, corresponding to twist $t=3$ at the amplitude level, for which  QCD factorization is not proven. In fact  an explicit computation of the $\rho_T$ electroproduction \cite{Mankiewicz:1999tt} at leading order shows  that the hard part has {\em end-point singularities} like
\beq
\label{end_point}
\int\limits_0^1 \frac{du}u \qquad {\rm and } \qquad 
\int\limits_0^1 \frac{du}{1-u}
\eq
occuring when the momentum fraction carried by the quark or the anti-quark vanishes.

\subsection{Some solutions to factorization breaking?}
\label{subSec:solutions}

In order to extend the factorization theorem at higher twist, as well as to improve the phenomenological description of hard exclusive processes at moderate values of the hard scale, several solutions have been 
proposed. 
%
%
%
One may  add contributions of \alert{3-parton DAs} \cite{Ball:1998sk,Ball:1998ff} for $\rho_T$ \cite{Anikin:2002uv,Anikin:2002wg} (of dominant twist equal 3 for $\rho_T$). 
This in fact does not solve the problem,  while reducing the level of
divergency, but is needed for consistency.


On top of the potential end-point singularities discussed above, phenomenologicaly the use of 
simple asymptotical DAs lead usually to a too small ERBL contribution in hard exclusive processes, a situation which is not improved by NLO corrections. It was suggested by Chernyak and Zhitnitsky
\cite{Chernyak:1983ej} to use DAs which would be mostly concentrated close to the end point, and not identical to the asymptotical DA, a solution which indeed improve very much the description of the data, for example of the pion form factor.
 However, since close to the end-point one may face theoretical inconsistencies when justifying the factorization, Li and Sterman \cite{Li:1992nu} then introduced
an improved collinear approximation (ICA). 
They suggested to keep a 
 transverse \alert{$\ell_\perp$} dependency in the $q$, $\bar{q}$ momenta.
Soft and collinear gluon exchange between the valence quarks are responsible for large double-logarithmic effects which exponentiate. 
The corresponding study is made easier when using the impact parameter space $b_\perp$ conjugated to $\ell_\perp\,,$ leading to the
{\aut Sudakov} factor
\beq
\label{SudakovFact}
\exp [-S(u, b, Q)   ]\,,
\eq
a factor already involved in previous studies of elastic hadron-hadron scattering at fixed angle \cite{Botts:1989kf}.
$S$ diverges when ${\stmath b_\perp \sim O(1/\Lambda_{QCD})}$ (large transverse separation, i.e. \structure{small transverse momenta}) or \structure{small fraction} 
${\stmath u \sim O(\Lambda_{QCD}/Q)}\,.$
This thus regularizes potential end-point singularities, even when using non asymptotical DAs. See Ref.~\cite{Musatov:1997pu} for a detailled
and pedagogical discussion in the case of the $\gamma \gamma^* \to \pi^0$ form factors. These Sudakov effects have been implemented outside of pure QCD processes,
 in particular for the study of semi-leptonic 
$B \to \pi$ 
 decay \cite{DescotesGenon:2001hm}.
In this ICA, a dependency of the hard part with respect to the partons transverse momenta is kept. This suggested Jakob and Kroll to keep such a dependency also inside the wave function of the produced meson. This was implemented in the form of a an ad-hoc non-perturbative gaussian ansatz \cite{Jakob:1993iw}
\beq
\label{GaussianDA} 
\exp [ -a^2 \, |k_\perp^2|/(u \bar{u}) ] \,,
\eq  
and other similar ans\"atze,
which give back the usual asymptotic DA $6 \, u \, \bar{u}$ when integrating over $k_\perp\,.$
These gaussian ans\"atze combined with the perturbative Sudakov resummation tail effect were then implemented 
 for various phenomenological studies like the pion form factor \cite{Jakob:1993iw}, the meson-photon form factor \cite{Jakob:1994hd,Kroll:1996jx}. The phenomenological description of the pion form factor is then improved, but is still below the data, even with the Chernyak and Zhitnitsky model. For other observables for which one really faces a end-point singularity, like the above example of $\rho_T$-electroproduction, the same approach seems to allow for a consistent treatment, and at least to interesting models \cite{Vanderhaeghen:1999xj,Goloskokov:2005sd,Goloskokov:2006hr,Goloskokov:2007nt} which can  describe the meson electroproduction data, in particular the HERA data at small-$x_{Bj}$.
We will in Sec.~\ref{SubSec:meson-HERA} that at small $x_{Bj}\,,$
relying on 
the $k_T-$factorization,  the off-mass-shellness
of the $t-$channel gluons can serve as a regulator, preventing from facing end-point singularities.

\section{A few applications}
\label{Sec:applications}

\subsection{Electroproduction of an exotic hybrid meson}
\label{SubSec:hybrid}

Using $\vec{J} = \vec{L}+\vec{S}$ and neglecting any spin-orbital interaction,
 $S$,  $L$ can be considered as additional quantum numbers to classify hadron states, with
\beqa
\label{mom}
\vec{J}^{\,2}=J(J+1)\,,\quad \vec{S}^{\,2}=S(S+1)\,, \quad \vec{L}^{\,2}=L(L+1),
\eqa
and \alert{$J=|L-S|\,, \cdots \,, L+S$}.
In the usual quark-model,
\structure{meson are $q \bar{q}$ bound states} with charge parity $C$ and space parity $P$ satisfying
\beq
\label{CP}
\alert{C=(-)^{L+S}} \quad {\rm and} \quad \alert{P=(-)^{L+1}}.
\eq
Thus the allowed quantum numbers are
\beq
\label{quantum-numbers}
\begin{array}{llll} 
S=0\,,& L=J ,\, &  J=0,\,1,\,2, ... \,: &J^{PC}= 0^{-+} {\footnotesize(\pi, \eta)}, \, 1^{+-} (h_1, b_1),\,2^{-+},\,3^{+-},\, ...\\
S=1\,,& L=0\,,& J=1 \,:&  J^{PC}= 1^{--} {\footnotesize(\rho, \omega, \phi)}\\
\phantom{S=1\,,} & L=1 \,, & J=0,\,1,\,2 \,:&
J^{PC}= 0^{++} {\footnotesize(f_0, a_0)}, \, 1^{++} {\footnotesize(f_1, a_1)},\,2^{++} {\footnotesize(f_2, a_2)} \\
\phantom{S=1\,,} & L=2\,,& J=1,\,2,\,3\,:&
J^{PC}= 1^{--}, \, 2^{--},\,3^{--}\\
...
\end{array}
\eq
which show that
the \alert{exotic} mesons with
 \alert{$J^{PC}= 0^{--}, \, 0^{+-},\, 1^{-+},\, \cdots$}
are forbidden.

We restrict ourselves to the light $1^{-+}$ exotic meson, denoted as $H$.
There are  several experimental candidates for $H$:
the $\pi_1(1400)$, seen at GAMS \cite{Alde:1988bv}, E852 \cite{Thompson:1997bs},
 Crystal Barrel \cite{Abele:1998gn,Abele:1999tf}, VES \cite{Dorofeev:2001xu},
%
%
%
%
%
%
%
%
%
%
%
 the $\pi_1(1600)$, seen at E852 \cite{Adams:1998ff,Ivanov:2001rv,Chung:2002pu,Kuhn:2004en,Lu:2004yn}, Crystal Barrel \cite{Baker:2003jh}, VES \cite{Beladidze:1993km,Khokhlov:2000tk,Dorofeev:2001xu,Amelin:2005ry}, most recently confirmed by  COMPASS \cite{:2009xt}, and 
%
%
%
%
%
%
the $\pi_1(2000)$ \cite{Kuhn:2004en,Lu:2004yn}.

 Based on the fact that an extra degree of freedom is required to describe these exotic quantum numbers \cite{Jaffe:1985qp,Bali:2000gf}, one possibility is to consider a tower of Fock states starting with
$|q \bar{q} g \rangle$ ($|q \bar{q} q \bar{q}  \rangle$ states may also be considered).  The natural question  is then to study the feasibility of producing exotic meson in  hard exclusive processes.  Based on the fact that such a Fock state is expected to be a higher twist component (of twist 3 when thinking
of the genuine twist 3 content of the usual $\rho$-meson), a strong $1/Q$ suppression was expected in hard electroproduction of $\pi_1$ with respect to $\rho$. It was shown in Refs.~\cite{Anikin:2004vc,Anikin:2004ja} that no suppression should be expected. This is based on the fact that the gluonic field operator does not need to appear explicitely in the local interpolating operator ${\cal O} (\Psi, \, \bar{\Psi}, \, A)$ creating the  $|q \bar{q} g \rangle$ state. Indeed, while the twist of such a typical operator $\bar{\Psi} \gamma^\mu G_{\mu \nu} \Psi$ is 4, leading to a $1/Q^2$ suppression, collinear approach describes 
hard exclusive processes in terms of \structure{non-local light-cone} operators, among which are the  \alert{twist 2 operator}
\beq
\label{non-local-operator}
\bar \psi(-z/2)\gamma_\mu [-z/2;z/2]
\psi(z/2)
\eq
where $[-z/2;z/2]$ is a Wilson line, necessary to fullfil gauge invariance (i.e. a ''color tube`` between $q$ and $\bar{q}$) which thus hides gluonic degrees of freedom: at twist 2 the needed gluon is there.

The $H$ DA is defined as (for longitudinal polarization)
\beq
\label{def-DA-H}
\langle H(p,0)| \bar \psi(-z/2)\gamma_\mu [-z/2;z/2]
\psi(z/2) | 0 \rangle_{\!\left|\!\tiny\begin{array}{lcc}z^2&\!\!\!\!\!=&\!\!\!\!0\\z+&\!\!\!\!\!=&\!\!\!\!0\\z_\perp&\!\!\!\!\!=&\!\!\!\!0\end{array}\right.}\hspace{-.1cm}=
i f_H M_H e^{(0)}_\mu\hspace{-.15cm}
\int\limits_0^1 \! dy \, e^{i(\bar y - y)p\cdot z/2} \phi^{H}_L(y)\,.
\eq
Inserting the $C$-parity operator gives an \alert{antisymmetric DA for $H^0$},
$\phi^{H}_L(y)=-\phi^{H}_L(1-y)$,  while the usual $\rho$ DA is symmetric.
The identification of quantum numbers can be performed when
expanding the operator in the l.h.s of Eq.~(\ref{def-DA-H}) in terms of \alert{local operators}
\begin{eqnarray}
\label{locdec}
&&\langle H(p,\lambda)| \bar\psi(-z/2) \gamma_{\mu}[-z/2;z/2] \psi(z/2)| 0\rangle
\nonumber \\
&&=\sum_{\stmath n}\frac{1}{n!}z_{\mu_1}..z_{\mu_n} \langle H(p,\lambda)|
\alert{\bar\psi(0) \gamma_{\mu}
\stackrel{\leftrightarrow}{D}_{\mu_1}..\stackrel{\leftrightarrow}{D}_{\mu_n}
\psi(0)}| 0\rangle ,
\nonumber
\end{eqnarray}
where $D_{\mu}$ is the usual covariant derivative and
$
\stackrel{\leftrightarrow} {D_{\mu}}=\frac{1}{2}(
\stackrel{\rightarrow}{D_{\mu}}-
\stackrel{\leftarrow}{D_{\mu}})\,.
$
The hybrid selects the odd-terms
\begin{eqnarray}
\label{locdec-hyb}
&&\langle H(p,\lambda)| \bar\psi(-z/2) \gamma_{\mu}[-z/2;z/2] \psi(z/2)| 0\rangle=
\nonumber \\
&&\sum_{\alert{n\, odd}}\frac{1}{n!}z_{\mu_1}..z_{\mu_n} \langle H(p,\lambda)|
{\bar\psi(0) \gamma_{\mu}
\stackrel{\leftrightarrow}{D}_{\mu_1}..\stackrel{\leftrightarrow}{D}_{\mu_n}
\psi(0)}| 0\rangle \,,
\nonumber
\end{eqnarray}
while the usual $\rho$-meson would select the even terms. The special case ${\stmath n=1}$ is just
\beq
\label{hybrid-n1}
{\cal R}_{\mu\nu}=\mbox{ S}_{(\mu \nu)}
\bar\psi(0)\gamma_{\mu}
\stackrel{\leftrightarrow}{D}_{\nu}\psi(0),
\eq
with $\mbox{S}_{(\mu\nu)}$ the symmetrization
operator
$\mbox{S}_{(\mu \nu)}T_{\mu \nu}=\frac{1}2(T_{\mu \nu}+T_{\nu
\mu})\,.$
The relation with the hybrid DA is now
\begin{eqnarray}
\label{emee}
\langle H(p,\lambda) | {\cal R}_{\mu\nu} |0\rangle=
\frac{1}{2}\,f_H \,M_H \,\mbox{\large S}_{(\mu \nu)}\,
e_{\mu}^{(\lambda)} \,p_{\nu} \int\limits_{0}^{1}dy (1-2y) \phi^{H}(y)\,.
\end{eqnarray}
The $C$- and $P$- parity are consistent since $C(R_{\mu \nu}) =+$ and 
$P(R_{k0}) =-$ (after going to rest-frame: $p_i=0$ and $e_0=0$).
The last step to control the order of magnitude is to
 fix $f_H$ (the analogue of $f_\rho$). It turns out that 
the operator ${\cal R}_{\mu\nu}$ is related to quark energy-momentum tensor $\Theta_{\mu\nu}:$
${\cal R}_{\mu\nu} = - i \, \Theta_{\mu\nu}\,$ which was studied based on
 QCD sum rules \cite{Balitsky:1982ps,Balitsky:1986hf}. Using the resonance  for $M \approx 1.4$ GeV
(the $\pi_1(1400)$) one gets
$\alert{f_H \approx 50 \,{\rm MeV}}$, to be compared with 
$f_\rho = 216\,$MeV.
This leads to the following rough estimate of ratios of electroproduction cross-sections
\beq
\label{ratio-electroproduction}
\alert{\frac{d\sigma^{H}(Q^2, x_B,  t )}{d\sigma^{\rho}(Q^2, x_B, t )}}
\approx \biggl( \frac{5 f_H}{3 f_\rho}\biggr)^2 \approx \alert{0.15}\,,
\eq
which does not change significantly \cite{Anikin:2004ja} when 
using {\em Double Distributions} \cite{Radyushkin:1997ki,Radyushkin:1998es} to model GPDs as well as when varying factorization and renormalization scales.

It turns out that the range around 1400 MeV is dominated by the 
 $a_2(1329)(2^{++})$ resonance, providing a possible playground for 
 interference effects between $H$ and $a_2$. This is possible  through the $\pi \eta$ channel, the presumable  main 
decay mode for the $\pi_1(1400)$ candidate. Based on models for the two $C=+$ and $C=-$  corresponding GDAs,  \structure{angular asymmetry} studies can be performed with respect to the
$\pi$ polar angle in the  $\pi\eta$ center-of-mass.

\begin{figure}
\psfrag{pf}[cc][cc]{$$}
\psfrag{Th}[cc][cc]{$H$}
\psfrag{Da}[cc][cc]{$DA$}
\psfrag{r}[cc][cc]{$\,\,\,H^0$}
\quad
\scalebox{.9}{$\begin{array}{ccccc}
\raisebox{-0.46 \totalheight}{\epsfig{file=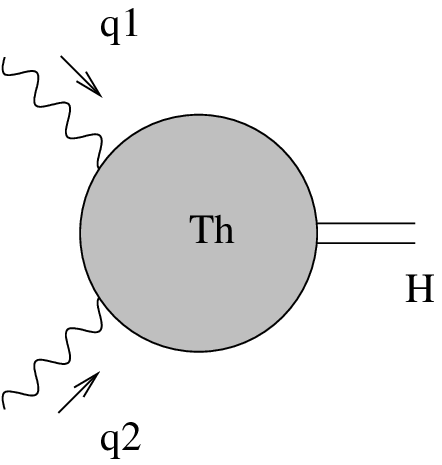,width=2cm}}
&=&
\raisebox{-0.46 \totalheight}{\epsfig{file=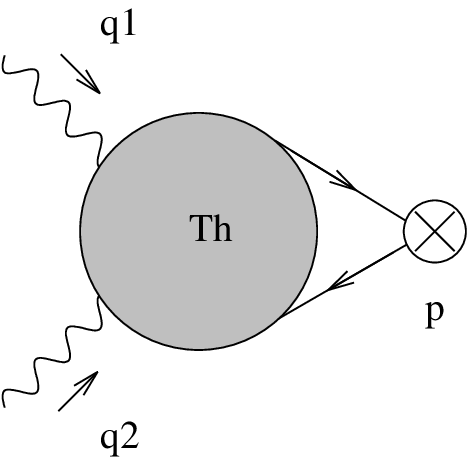,width=2cm}}
&&
\raisebox{-0.4\totalheight}{\epsfig{file=DA.eps,width=1.5cm}}
\end{array}$}
\psfrag{p}[cc][cc]{$$}
\psfrag{p}[cc][cc]{$$}
\psfrag{q1}[cc][cc]{$\gamma^*$}
\psfrag{q2}[cc][cc]{$\gamma$}
\psfrag{Th}[cc][cc]{$H$}
\psfrag{n}[cc][cc]{$$}
\quad 
{\scalebox{.75}
{ \quad
$\begin{array}{ccccc}
\raisebox{-0.46 \totalheight}{\epsfig{file=THT.eps,width=2cm}}
&=&
\raisebox{-0.46 \totalheight}{\epsfig{file=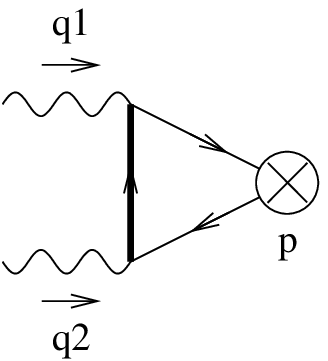,width=2cm}}
&+&
\raisebox{-0.46\totalheight}{\epsfig{file=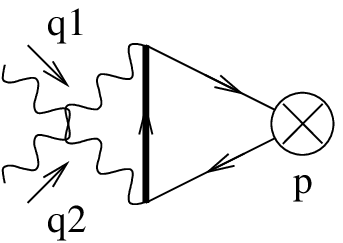,width=2cm}}
\end{array}
$}}
\caption{Left: Factorization of the $\gamma^* \gamma \to H$ process. Right: Hard part at leading order.}
\label{Fig:gamma-gamma*-H}
\end{figure}

Hybrid could be also copiously produced in $\gamma^* \gamma$ channel, i.e. at $e^+ e^-$ colliders \alert{with one tagged out-going electron}.  This can be described in a hard factorization framework, as illustrated in Fig.~\ref{Fig:gamma-gamma*-H}.  
%
\psfrag{p}[cc][cc]{$$}
\psfrag{q1}[cc][cc]{$\gamma^*$}
\psfrag{q2}[cc][cc]{$\gamma$}
\psfrag{Th}[cc][cc]{$$}
\psfrag{e1}[cc][cc]{$\!\!\!e^\pm$}
\psfrag{e2}[cc][cc]{$\!\!\!e^\pm$}
\psfrag{H}[cc][cc]{$H^0$}
The basic result obtained in this framework is that the production amplitude 
for a hybrid state $M^{\gamma^* \gamma  \to \pi_1} $ scales in $Q^2$ in the same way as the one for the
 "non-exotic'' $\pi^0$ production. 
One also obtains an estimate for the ratio of squared matrix elements of scattering amplitude 
for a hybrid state $M^{\gamma^* \gamma  \to \pi_1} $ versus a "non-exotic'' $\pi^0$ production  
\beq
\label{ratio}
\frac{|M^{\gamma^* \gamma  \to \pi_1}|^2}{|M^{\gamma^* \gamma  \to \pi^0}|^2} \simeq 20 \% \,.
\nonumber
\eq
Based on BaBar counting rates of $\gamma^* \gamma \to \eta'$ up to $Q^2=30$ GeV$^2$, one expect visible
counting rates for  $\gamma^* \gamma \to \pi_1$\,.
If the state does not appear as a bump in the mass distribution, one may look for interference effects with the background
opening the possibility to 
enhance the hybrid signal \cite{Anikin:2006du}.

\subsection{Spin transversity in the nucleon}
\label{SubSec:transversity}

The transverse spin content of the proton is an observable which is non-diagonal with respect to helicity. Indeed,  
\beq
\label{transversity}
\begin{tabular}{ccc}$| \uparrow \rangle_{(x)}  $&$\sim$& $| \rightarrow\rangle +| \leftarrow\rangle$\\
$| \downarrow \rangle_{(x)}  $&$\sim$& $| \rightarrow\rangle -| \leftarrow\rangle\,.$\\
 spin along  $x$ & & helicity state
\end{tabular}
\eq
An observable sensitive to helicity spin flip
gives thus access to the transversity
\alert{$\Delta_T q(x)$},
which is very badly known.
The transversity GPDs themselves are completely unknown.
Chirality $\pm$ is defined by
\beq
\label{def-chirality} 
q_\pm(z) \equiv \frac{1}{2}(1\pm \gamma^5)q(z)\, \, \hbox{  with  }\, \, q(z)= q_+(z) + q_-(z)
\,.
\eq
A chiral-\alert{even} quantity {\em conserves} chirality, like
$\bar q_{\pm}(z) \gamma^\mu q_\pm(-z)$ and  $\bar q_\pm(z) \gamma^\mu \gamma^5q_\pm(-z)\,,$
while 
a chiral-\alert{odd} operator {\em reverses} chirality, like
$\bar q_{\pm}(z)\cdot 1\cdot q_\mp(-z),\;\;\;
\bar q_{\pm}(z)\cdot \gamma^5\cdot q_\mp(-z)
$ and $\bar q_\pm(z) [\gamma^\mu ,\gamma^\nu] q_\mp(-z)\,.$
For a 
 massless (anti)particle,  chirality = (-)helicity.
\alert{Transversity is thus a chiral-odd quantity}.
Now, since QCD and QED are chiral even (neglecting mass effects), the observable we are looking for should have the form
$\, {\cal A}\sim (\mbox{Ch.-\alert{odd}})_1\otimes (\mbox{Ch.-\alert{odd}})_2\,.$

\begin{figure}[h]
\psfrag{g}[cc][lc]{\raisebox{0.3 \totalheight}{\footnotesize$\gamma^*$}}
\psfrag{p}[cB][lc]{\footnotesize$N$}
\psfrag{pp}[lB][lc]{\footnotesize$N'$}
\psfrag{ro}[ct][ct]{\footnotesize$\,  \stmath{\rho_T}$}
\psfrag{H}[cc][cc]{GPD}
\vspace{.5cm}
\scalebox{1}{\centerline{\begin{tabular}{ccc}
\hspace{0cm}\raisebox{0 \totalheight}{\includegraphics[height=2.6cm]{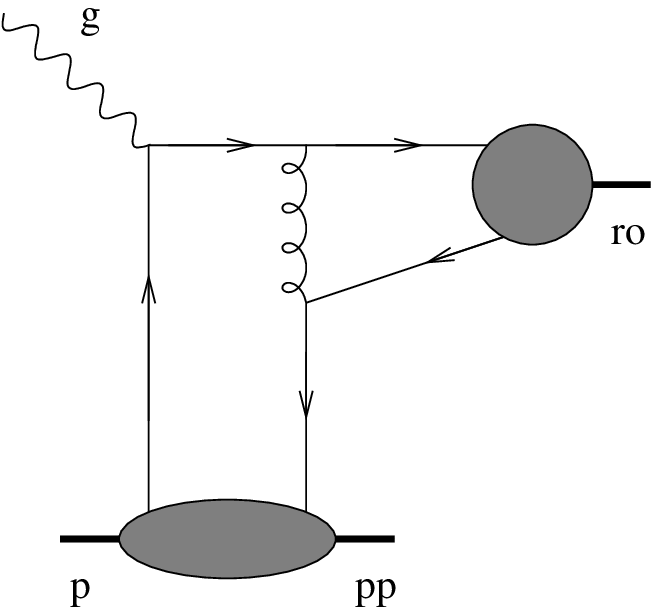}} 
& \qquad&\hspace{0cm}\raisebox{0 \totalheight}{\includegraphics[height=2.6cm]{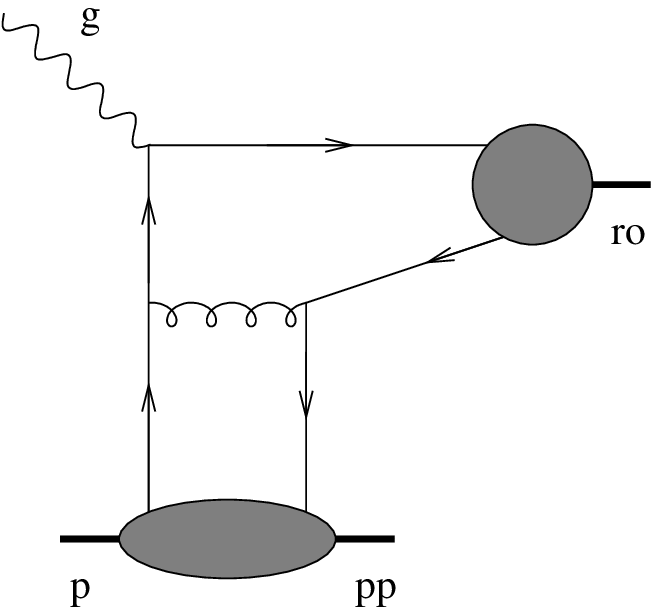}}  \end{tabular}}}
\caption{The two diagrams contributing at twist 2 to $\gamma^* \, N\to \rho_T \, N'$.}
\label{Fig:transversity}
\end{figure}

\begin{figure}[h]
\vspace{.2cm}
\begin{center}
\begin{tabular}{ccc}
\psfrag{z}{\hspace{-0.1cm}\footnotesize $z$ }
\psfrag{zb}{\raisebox{0cm}{\hspace{-0.1cm} \footnotesize$\bar{z}$} }
\psfrag{gamma}{\raisebox{+.1cm}{\footnotesize $\,\gamma$} }
\psfrag{pi}{\footnotesize$\!\pi$}
\psfrag{rho}{\footnotesize$\,\rho$}
\psfrag{TH}{\raisebox{-.05cm}{\hspace{-0.1cm}\footnotesize $T_H$}}
\psfrag{tp}{\raisebox{.3cm}{\footnotesize $t'$}}
\psfrag{s}{\hspace{0.4cm}\footnotesize$s$ }
\psfrag{Phi}{}
\hspace{-0.2cm}
\scalebox{.9}{\raisebox{.6cm}{\includegraphics[width=3.7cm]{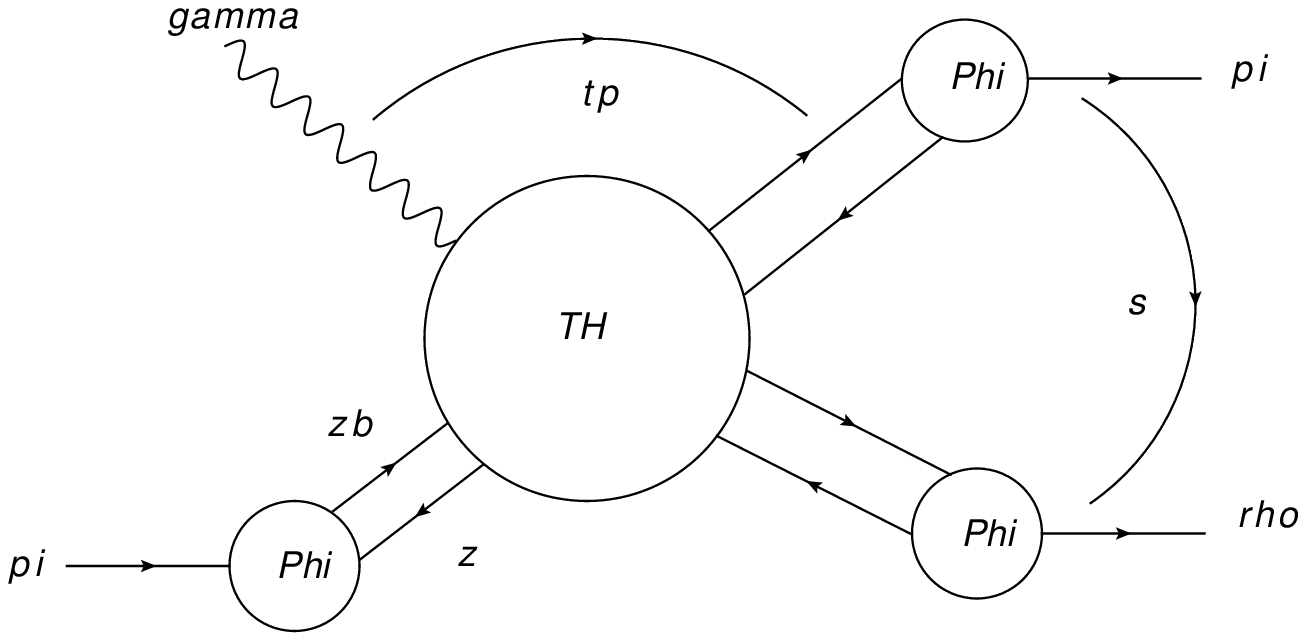}}}
& 
\psfrag{Phi}{}
\psfrag{gamma}{\raisebox{+.1cm}{\footnotesize $\,\gamma$} }
\psfrag{tp}{\raisebox{.3cm}{\footnotesize $t'$}}
\psfrag{piplus}{\footnotesize$\,\pi^+ $ \begin{tabular}{l} chiral-\alert{even} \\ twist 2 DA \end{tabular}}
\psfrag{rhoT}{\footnotesize$\,\rho^0_T$ \begin{tabular}{l} chiral-\alert{odd} \\ twist 2 DA
\end{tabular}}
\psfrag{M}{\hspace{-0.15cm} \footnotesize $M^2_{\pi \rho}$ }
\psfrag{x1}{\raisebox{-.1cm}{\hspace{-0.6cm} \footnotesize $x+\xi $  }}
\psfrag{x2}{\raisebox{-.1cm}{\hspace{-0.1cm} \footnotesize  $x-\xi $ }}
\psfrag{N}{ \hspace{-0.4cm}\footnotesize $p$}
\psfrag{GPD}{\raisebox{-.1cm}{\footnotesize \hspace{-0.4cm}  $GPDs$}}
\psfrag{Np}{\footnotesize$n$}
\psfrag{t}{ \raisebox{-.1cm}{\footnotesize \hspace{-0.3cm}   $t \ll M^2_{\pi \rho}$  \hspace{.1cm}\raisebox{-.1cm}{chiral-\alert{odd} twist 2 GPD}}}
\hspace{0.3cm}
\scalebox{.9}{\includegraphics[width=3.5cm]{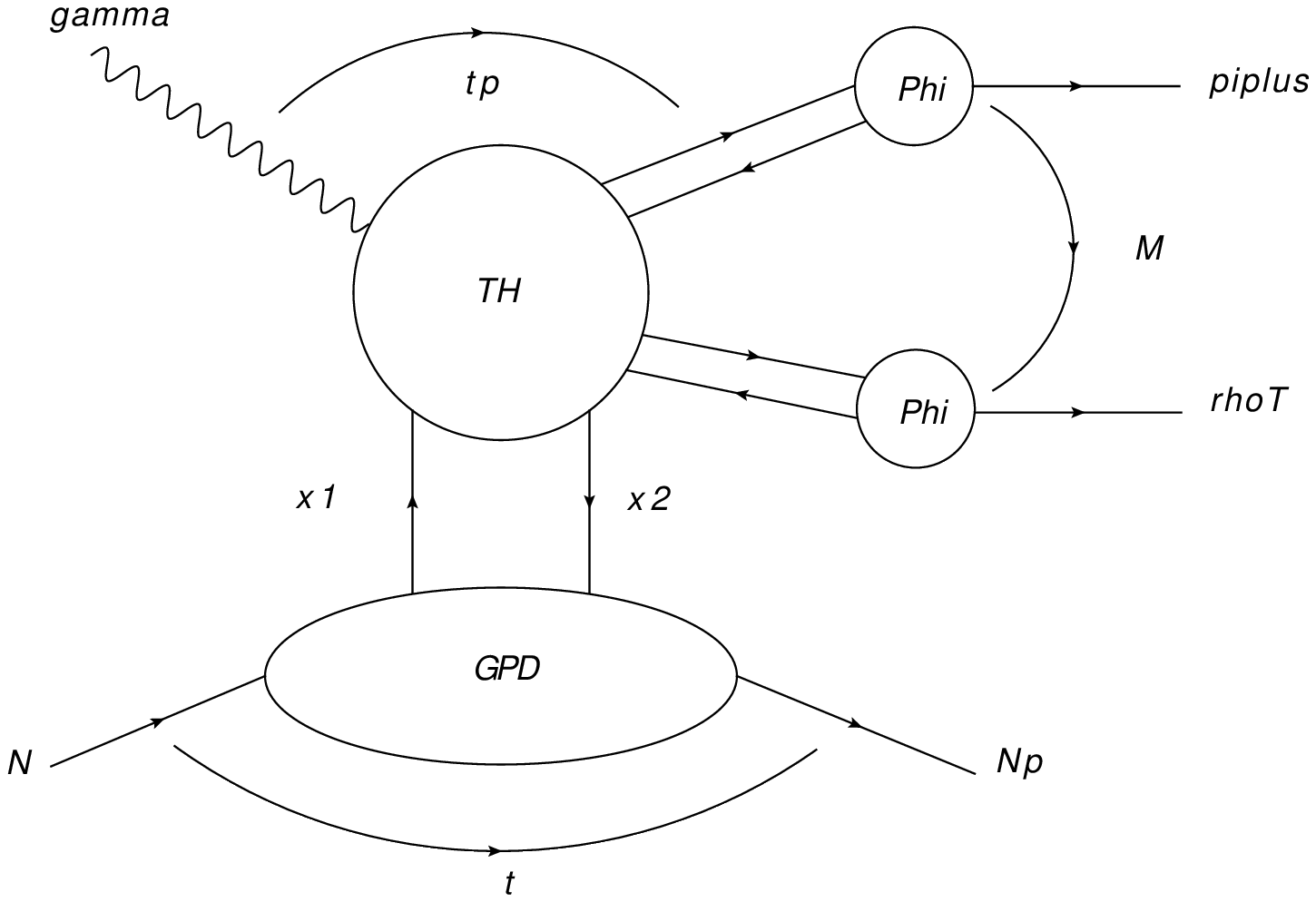}}
&
\psfrag{fpi}{}
\psfrag{fro}{}
\psfrag{z}{}
\psfrag{zb}{}
\psfrag{v}{}
\psfrag{vb}{}
\psfrag{gamma}{\footnotesize$\,\gamma$}
\psfrag{pi}{\footnotesize$\,\pi^+$}
\psfrag{rho}{\footnotesize$\,\rho^0_T$}
\psfrag{N}{\hspace{-.1cm}\footnotesize$p$}
\psfrag{Np}{\footnotesize$n$}
\psfrag{H}{\raisebox{-.06cm}{\tiny\hspace{0cm} $H^{ud}_T$}}
\psfrag{p1}{}
\psfrag{p2}{}
\psfrag{p1p}{}
\psfrag{p2p}{}
\psfrag{q}{}
\psfrag{ppi}{}
\psfrag{prho}{}
\hspace{-1.15cm}
\scalebox{1}{\raisebox{0cm}{\hspace{3cm}\includegraphics[width=2.3cm]{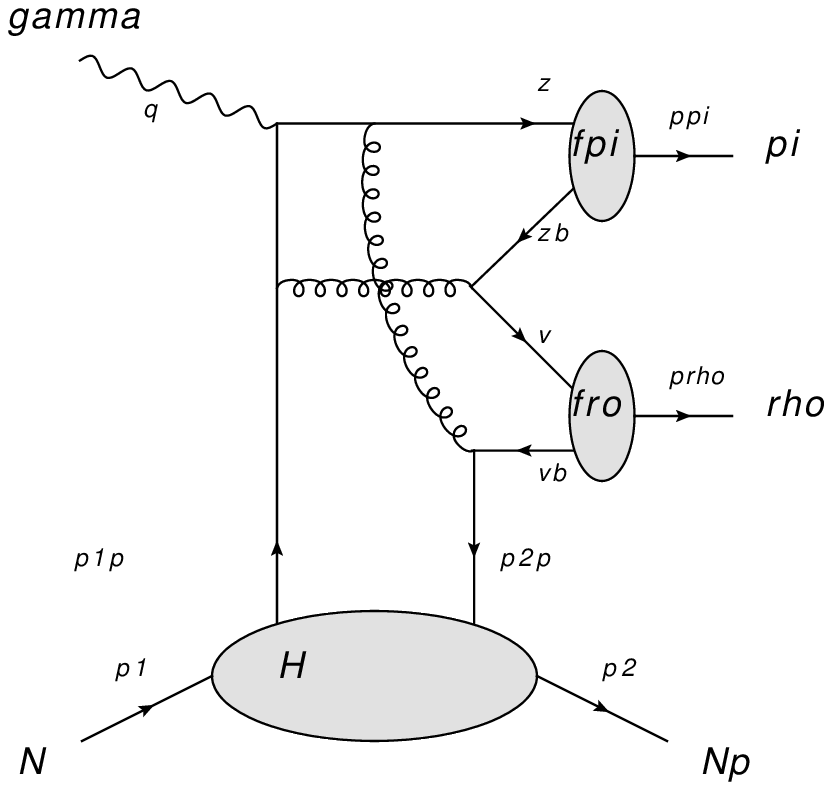}}}
\end{tabular}
\end{center}
\caption{Left: Factorization of the amplitude for the process $\gamma + \pi \rightarrow \pi + \rho $ at large $s$ and fixed angle (i.e. fixed ratio $t'/s$); Center: replacing one DA by a GPD leads to the factorization of the amplitude  for $\gamma \, p \rightarrow \pi^+ \, \rho^0_T \, n$ at large $M_{\pi\rho}^2$. Right: An example of non-vanishing diagram contributing to $\gamma p \to \pi^+ \rho^0_T n$.}
\label{Fig:transversity-process}
\end{figure}

The dominant DA for $\rho_T$ is of twist 2 and chiral-odd
($[\gamma^{\mu},\gamma^\nu]$ coupling). 
Unfortunately, the scattering amplitude of the process  $\gamma^* \, N \to \rho_T \, N'$ is zero at twist 2. Indeed, at Born order, the two diagrams shown in Fig.~\ref{Fig:transversity} vanish  \cite{Diehl:1998pd}, due to
$\gamma^\alpha[\gamma^{\mu},\gamma^\nu]\gamma_\alpha=0.$
This is true at any order in perturbation theory \cite{Collins:1999un}, since this would require a transfer of 2 units of helicity from the proton.
 This vanishing is true only a twist 2. A possible way out is to consider higher twist contributions,
which do not vanish \cite{Ahmad:2008hp,Goloskokov:2009ia}.
However processes involving twist 3 DAs may face problems with factorization
(end-point singularities: see later).
%
%
%
The process $\gamma p \to \pi^+ \rho^0_T n$ gives access to transversity at twist 2. The factorization picture of this process is similar to the
factorization \`a la  Brodsky Lepage of $\gamma + \pi \rightarrow \pi + \rho $ at large $s$ and fixed angle (i.e. fixed ratio $t'/s, u'/s$), as shown in Fig.~\ref{Fig:transversity-process} (left).
This justifies the factorization of the amplitude  for $\gamma p \to \pi^+ \rho^0_T n$ \structure{at large $M_{\pi\rho}^2$}, as shown in Fig.~\ref{Fig:transversity-process} (center).
A typical non-vanishing diagram is shown in Fig.~\ref{Fig:transversity-process} (right)
At large $s$, with $\pom$omeron exchange, a similar study was proposed earlier \cite{Ivanov:2002jj,Enberg:2006he}.
All these processes with a 3 body final state can give access to all GPDs:
\structure{$M_{\pi\rho}^2$ plays the role of the $\gamma^*$ virtuality of usual DVCS (here in the time-like domain)} and could be studied at {\aut JLab and COMPASS}.




\section{Hard exclusive processes in the perturbative Regge limit}
\label{Sec:Regge}

\subsection{Theorical motivations}
\label{SubSec:motivations}

Consider the diffusion of two hadrons $h_1$ and $h_2$, in the special limit 
where 
\beq
\label{regge-kinematics}
\alert{\sqrt{s}} \,
\gg {\rm \, other \ scales \ (masses, transfered \ momenta, virtualities ...)}
\gg \,\Lambda_{QCD}\,.
\eq
In this limit, typical large logarithms like ${\stmath \alpha_s \, \ln s \sim 1}$
arise, and should be resummed. The dominant sub-series, in the called LLx approximation,
\beq
\begin{array}{ccccc} {\cal A}=\raisebox{-0.35 \totalheight}{\epsfig{file=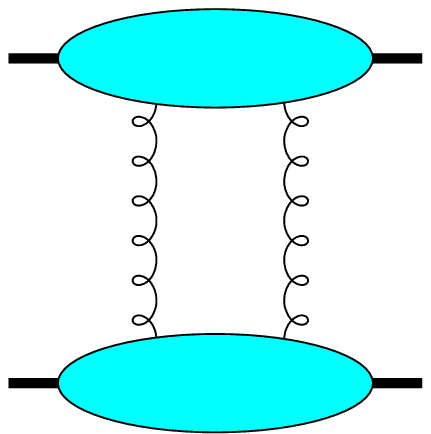,width=1.1cm}} \hspace{0cm}&+&  \left(\,   \raisebox{-0.35 \totalheight}{\epsfig{file=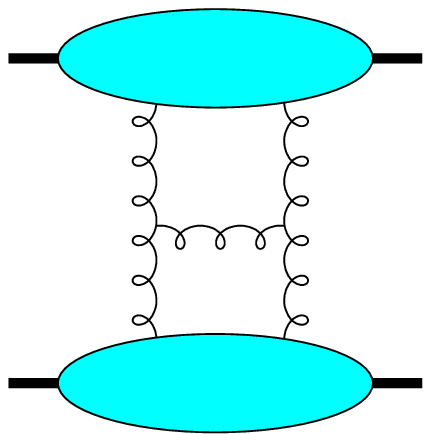,width=1.1cm}} \hspace{0cm}+  \raisebox{-0.35 \totalheight}{\epsfig{file=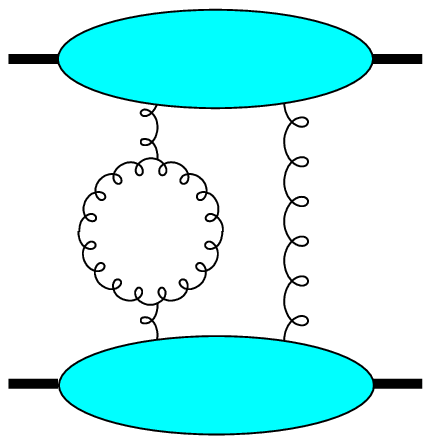,width=1.1cm}} + \cdots \right ) &+& \left(\raisebox{-0.35 \totalheight}{\epsfig{file=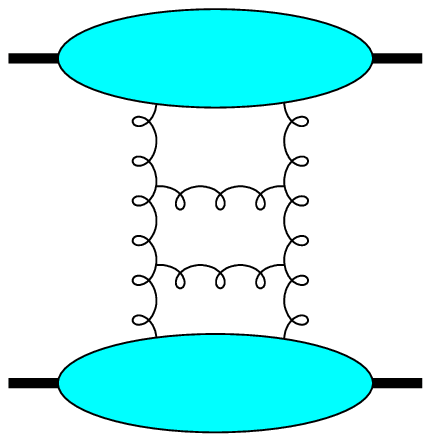,width=1.1cm}}+ \cdots \right) + \cdots\\
\qquad \sim s& & \sim s \, (\alpha_s \ln s) & & \hspace{-.8cm} \sim s \, (\alpha_s \, \ln s)^2  
   \end{array}
\label{Regge-diagrams}
\eq
leads, using the optical theorem,  to the total-cross section
\beq
\label{total-cross-section-LLx}
\alert{\sigma_{tot}^{h_1\, h_2 \to tout}} =\frac{1}s Im \,{\cal A} \sim \alert{ s^{\alpha_\pom(0) -1}}\,,
\eq
with $\alpha_\pom(0)-1= C \, \alpha_s \ (\alert{C >0})\,.$
This is the so-called 
 BFKL \alert{$\pom$omeron} (Balitsky, Fadin, Kuraev, Lipatov) \cite{Fadin:1975cb,Kuraev:1976ge,Kuraev:1977fs,Balitsky:1978ic}.
This result violates  QCD $S$ matrix unitarity
which states that $S \, S^\dagger=S^\dagger \, S=1$ ( i.e. \ \alert{$\sum Prob. =1$}). The question is thus 
until when this result could be applicable, and how to improve it. Phenomenologically, a longstanding question is 
how to test this dynamics experimentally, in particular based on exclusive processes.


\subsection{$k_T$ factorization}
\label{SubSec:kT-factorization}

\begin{figure}
\psfrag{g1}[cc][cc]{$\gamma^*(q_1)$}
\psfrag{g2}[cc][cc]{$\gamma^*(q_2)$}
\psfrag{p1}[cc][cc]{\raisebox{.3cm}{$\,\rho(p_1)$}}
\psfrag{p2}[cc][cc]{\raisebox{.3cm}{$\,\rho(p_2)$}}

\psfrag{l1}[cc][cc]{$l_1$}
\psfrag{l1p}[cc][cc]{$-\tilde{l}_1$}
\psfrag{l2}[cc][cc]{$l_2$}
\psfrag{l2p}[cc][cc]{$-\tilde{l}_2$}
\psfrag{ai}[cc][cc]{$\stmath \beta^{\,\nearrow}$}
\psfrag{bd}[cc][cc]{$\stmath \alpha_{\,\searrow}$}
\psfrag{k}[cc][cc]{$k$}
\psfrag{rmk}[lc][cc]{$\hspace{-.5cm}r-k \hspace{3cm} \int d^2 k_\perp$}
\psfrag{oa}[cc][cc]{${}\quad {\tiny \stmath \alpha \ll \alpha_{\rm quarks}}$}
\psfrag{ou}[ll][ll]{
\hspace{1cm} $\Rightarrow$ set $\alpha=0$ and  $\int d\beta$ }
\psfrag{ob}[cc][cc]{\raisebox{-1.5
    \totalheight}{$\quad {\tiny \stmath\beta \ll \beta_{\rm quarks}}$}}
\psfrag{od}[ll][ll]{
\hspace{1cm} $\Rightarrow$ set $\beta=0$ and  $\int d\alpha$}
\begin{center}
\hspace{-2cm}\epsfig{file=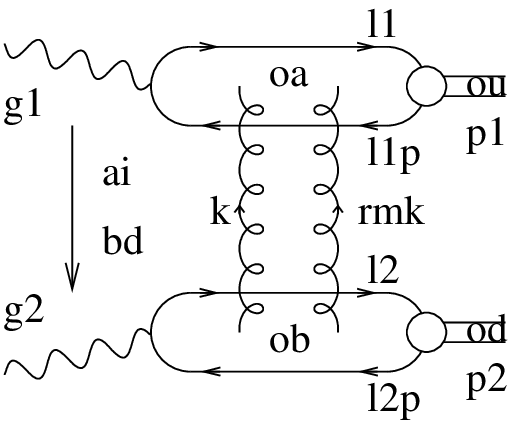,width=\wid}
\end{center}
\label{Fig:kt-factorization}
\end{figure}

Let us consider $\gamma^* \, \gamma^* \to \rho \, \rho$ scattering as an example.
Using the {\aut Sudakov} decomposition (\ref{expansion-Sudakov})
where the two outgoing mesons flies along $p_1$ and $p_2$, and expanding each loop momentum 
integration according to
\beq
\label{loop-k}
d^4k= \frac{s}{2} \, d \alpha \, d\beta \, d^2k_\perp
\eq
the dominant contribution for the amplitude, which scales like $s$ in the two gluons approximation,  is obtained in the approximation where the above and below gluon emissions are treated eikonally,
with $\alpha \ll \alpha_{\rm quarks}$ (above) and $\beta \ll \beta_{\rm quarks}$ (below). The amplitude is dominated by the exchange  of the $t-$channel gluons with \alert{non-sense} polarizations ($\varepsilon_{\alert{NS}}^{up}=\frac{2}{\sqrt{s}} \, p_2$, $\varepsilon_{\alert{NS}}^{down}=\frac{2}{\sqrt{s}}\,  p_1$). This 	
approximation is illustrated in Fig.~\ref{Fig:kt-factorization}
This leads to the impact representation for the amplitude, in the two-gluon approximation\footnote{Underlined letters denote euclidean two-dimensional vectors.},
\beq
{\cal M} = is\;\int\;\frac{d^2\,\kb}{(2\pi)^2\kb^2\,(\rb -\kb)^2}
\Phi^{\gamma^*(q_1) \to \rho(p^\rho_1)}(\kb,\rb -\kb)\;
{\Phi}^{\gamma^*(q_2) \to \rho(p^\rho_2)}(-\kb,-\rb +\kb)
\label{M-kt}
\eq
where
$\alert{{\Phi}^{\gamma^*(q_1) \to \rho(p^\rho_1)}}$ are the   $\gamma^*_{L,T}(q) g(k_1) \to \rho_{L,T}\, g(k_2)\,$. 
The LLx approximation is obtained when replacing the two gluon exchange by the BFKL ladder,  thus changing the first term  under the integration in Eq.~(\ref{M-kt}) by the BFKL Green function.

Note that the two $t-$channel gluons are off-shell, in contrast with usual collinear factorization.
Since probes are color neutral, QCD gauge invariance implies that
their impact factor should vanish  when $\kb \to 0$ or $\rb-\kb \to 0$. 

\subsection{Meson production at HERA}
\label{SubSec:meson-HERA}

Diffractive meson production at
{\aut HERA}, the first and single $e^{\pm}p$ \structure{collider}, running from 1992 until 2007, is a typical application of the above tool.
The ''easy'' case (from factorization point of view) is  $J/\Psi$ production: since $u \sim 1/2$ based on the non-relativistic limit for bound state of massive quarks \cite{Chernyak:1983ej}, one avoids possible end-point singularities \cite{Ryskin:1992ui,Martin:1996bp,Martin:1997sh,Martin:1999wb,Enberg:2002zy,Ivanov:2004vd}.
%
At large $t$ (providing the hard scale), light meson diffractive photoproduction $\gamma+p\to \rho_{L,T}+X$ (with a rapidity gap between the meson and the proton remnants) was studied at LLx
based on $k_T$-factorization \cite{Ivanov:1995nf,Forshaw:1995ax,Bartels:1996fs,Forshaw:2001pf}, taking into account a possible chiral-odd coupling of the photon \cite{Ivanov:2000uq,Enberg:2003jw,Poludniowski:2003yk}. In these approaches,
{\aut H1 and ZEUS} data seems to favor {\aut BFKL} 
	 but end-point singularities for $\rho_T$ are regularized   with a quark mass $m=m_\rho/2$
	while 
the spin density matrix is badly described.
	
\begin{figure}[h]
\psfrag{t}{}
\psfrag{M}{}
\centerline{\includegraphics[width=3cm]{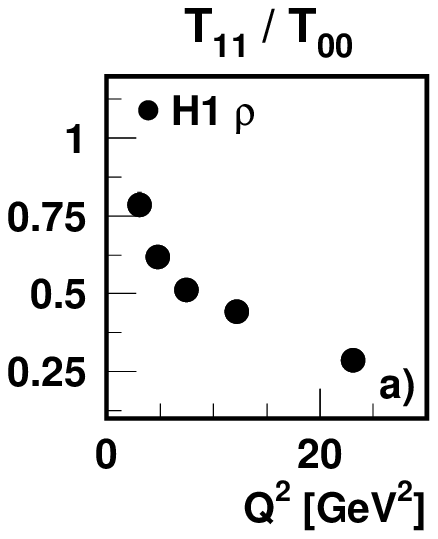} \hspace{.05cm} \includegraphics[width=3cm]{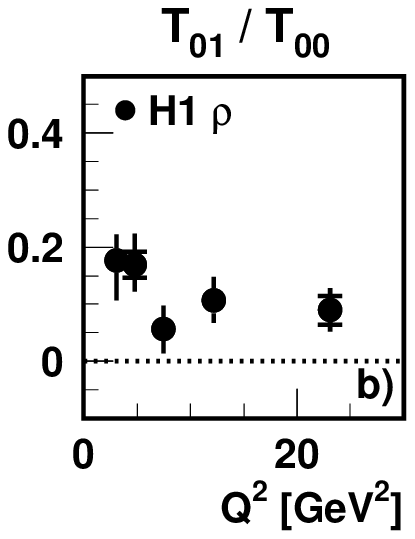} \hspace{.05cm} \includegraphics[width=3cm]{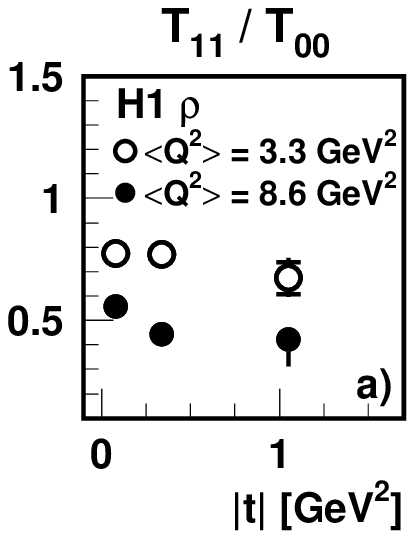} \hspace{.05cm} \includegraphics[width=3cm]{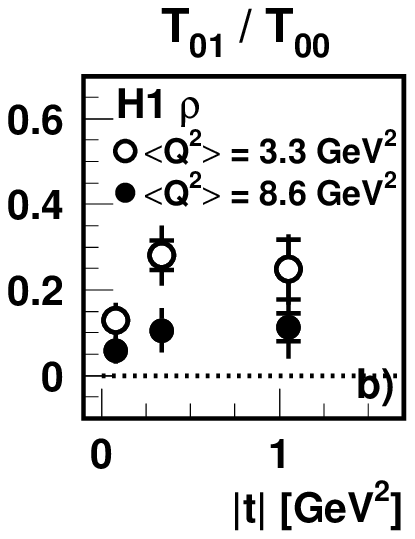}}
\caption{Left: Ratio $T_{11}/T_{00}$ (a) and $T_{01}/T_{00}$ (b) as a function of $|t|$. Right (a) and (b): same ratios as a function of $Q^2$, as measured by H1 for $\gamma^*_{L,T}+p\to \rho_{L,T}+p$. Figures from \cite{Aaron:2009xp}.}
\label{Fig:T-H1}
\end{figure}

 Exclusive electroproduction of vector meson  
${\stmath \gamma^*_{L,T}+p\to \rho_{L,T}+p}$ was studied in \linebreak Ref.~\cite{Ivanov:1998gk}
and a hierarchy for the helicity amplitudes $T_{\lambda_1 \lambda_2}$  of the process ($\lambda_1 = 0,+1,-1$ is the photon helicity and $\lambda_2 = 0,1,-1$ is the vector-meson helicity) was obtained,
modifying the pure SCHC according to
\beq
\label{SCHC}
T_{00} \, > \,  T_{11}  \, > \, T_{10}  \, > \, T_{01}  \, > \, T_{1-1}\,.
\eq
The recent HERA data \cite{Chekanov:2007zr,Aaron:2009xp} are in agreement with the above hierarchy, as illustrated in Fig.~\ref{Fig:T-H1}
for the ratios $T_{11}/T_{00}$ and $T_{01}/T_{00}$, the two left panel showing in particular the twist 2 dominance of the amplitude $T_{00}$ with respect to the twist 3 dominated  amplitudes $T_{11}$ and $T_{01}$. 
A similar approach to $k_T$-factorization, based on the so-called dipole model in transverse coordinate space \cite{Mueller:1989st,Nikolaev:1990ja}, has been developped \cite{Nemchik:1996cw} and applied to HERA data \cite{Forshaw:2003ki,Kowalski:2006hc,Forshaw:2010py,Forshaw:2011yj}. 
Besides,  it turns out that the data can also be well described by a GPD like evolution, based on  ICA for the coupling with the meson DA with a gaussian ansatz for the meson wave function combined with Sudakov resummation effects
\cite{Goloskokov:2005sd,Goloskokov:2006hr,Goloskokov:2007nt}. There is however no complete description of this process starting from first principle.

\begin{figure}
\psfrag{rho}[cc][cc]{$\rho$}
\psfrag{k}[cc][cc]{}
\psfrag{rmk}[cc][cc]{}
\psfrag{l}[cc][cc]{$\ell$}
\psfrag{q}[cc][cc]{}
\psfrag{lm}[cc][cc]{}
\psfrag{H}[cc][cc]{$ H_{q \bar{q}}$}
\psfrag{S}[cc][cc]{$ \Phi_{q \bar{q}}$}

\scalebox{.65}
{\begin{tabular}{ccccc}
\hspace{.5cm}\epsfig{file=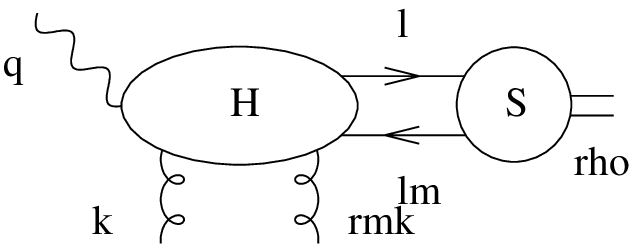,width=4.5cm}
& 
\hspace{-.2cm}\raisebox{.9cm}{$\longrightarrow $} 
&
\psfrag{lm}[cc][cc]{\raisebox{.2cm}{$\quad \,\,\, \, \, \Gamma \ \,\, \Gamma$}}
\hspace{-.4cm}\epsfig{file=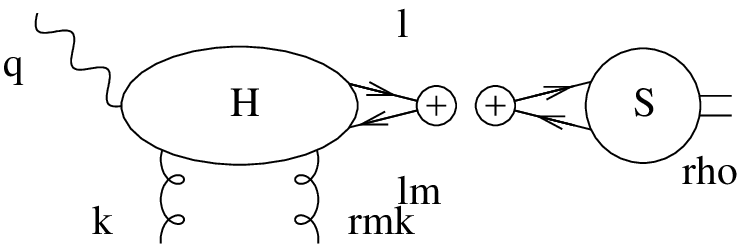,width=5.5cm}
&\hspace{-.2cm}\raisebox{.9cm}{$+$}
&
\psfrag{H}[cc][cc]{{$H^\perp_{q \bar{q}}$}}
\psfrag{S}[cc][cc]{\scalebox{1}{$\Phi^\perp_{ q \bar{q}}$}}
\psfrag{lm}[cc][cc]{\raisebox{.2cm}{$\quad \,\,\, \, \, \Gamma \ \,\, \Gamma$}}
\hspace{-.6cm}
\epsfig{file=FiertzHSqq_rhofact.eps,width=5.5cm}
\end{tabular}}


\psfrag{rho}[cc][cc]{$\rho$}
\psfrag{k}[cc][cc]{}
\psfrag{rmk}[cc][cc]{}
\psfrag{l}[cc][cc]{}
\psfrag{q}[cc][cc]{}
 \psfrag{Hg}[cc][cc]{$H_{q \bar{q}g}$}
 \psfrag{Sg}[cc][cc]{$\!\Phi_{q \bar{q}g}$}
 \psfrag{lm}[cc][cc]{}
 \psfrag{H}[cc][cc]{$H_{q \bar{q}g}$}
 \psfrag{S}[cc][cc]{$ \Phi_{q \bar{q}}$}
 \psfrag{S}[cc][cc]{$\Phi_{ q \bar{q}g}$}
 \scalebox{.65}{\begin{tabular}{ccc}
 \hspace{0cm}\epsfig{file=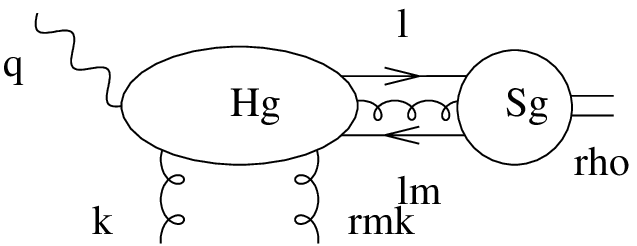,width=5cm}
&
\quad \raisebox{1.1cm}{$\longrightarrow $} \quad
&
\psfrag{lm}[cc][cc]{\raisebox{.2cm}{$\quad \ \, \,\,\, \, \, \, \Gamma \ \ \, \, \Gamma$}}
 \raisebox{.1cm}{\epsfig{file=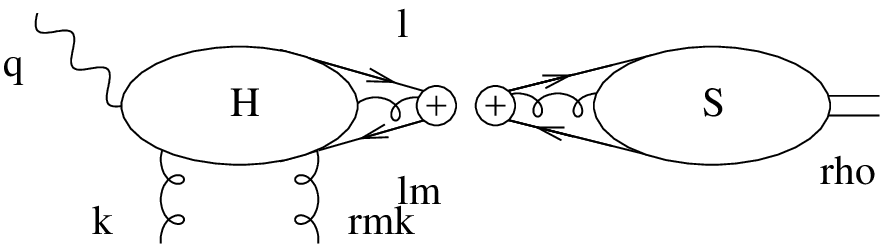,width=7cm}}
 \end{tabular}}
\caption{Impact factor factorization up to twist 3.}
\label{Fig:impact-factor-factorization}
\end{figure}

The light-cone collinear factorization has been developped in order to deal with exclusive processes beyond leading twist \cite{Anikin:2000em,Anikin:2001ge,Anikin:2002wg,Anikin:2002uv}, inspired by the inclusive case \cite{Efremov:1981sh,Shuryak:1981kj,Shuryak:1981pi,Ellis:1982cd,Efremov:1983eb,Teryaev:1995um,Radyushkin:2001fc}.
Recently, a new  self-consistent and very efficient
extension has been carried on
  at a full twist 3 level \cite{Anikin:2009hk,Anikin:2009bf}, illustrated below for the $\gamma^*_T \to \rho_T$ impact factor at twist 3.
 It is a non-covariant technique in axial gauge based on the parametrization of matrix element along a \structure{light-like prefered direction} $z = \lambda \, \alert{n}$ ($n=2 \, p_2/s$).
Using notations of Fig.~\ref{Fig:factorization-DA}, the pure twist 2 collinear approximation means  $l_\mu =\alert{u \,p_\mu}$, which we should now extend. A
{\aut Sudakov} expansion in the basis $p \sim p_\rho, \,n$ ($p^2=n^2=0$ and $p \cdot n =1$) is made, with the scaling indicated below each term 
\beq
\label{Sudakov-scaling}
\begin{array}{cccccccc}
l_\mu &=&\alert{u \,p_\mu}  &+& {\twist3 l^\perp_\mu} &+& (l\cdot p)\, n_\mu ,&
\quad \alert{u}=l\cdot n \\
\\
 & & {\stmath ~1} & & { \stmath ~1/Q} & & {\stmath ~1/Q^2}\,.
\end{array}
\eq
We now {\aut Taylor} expand the \alert{hard} part $H(\ell)$ along the collinear direction \alert{$p$}
\beq
\label{Taylor-hard}
H(\ell) = H(\alert{u p}) + \frac{\partial H(\ell)}{\partial \ell_\alpha} \biggl|_{\ell=u p}\biggr. \, (\ell-u \,p)_\alpha + \ldots
 \quad \text{with} \,\,\,\, (\ell-u\, p)_\alpha \approx{\twist3 \ell^\perp_\alpha} \,.
\eq
%
%
Fourier transform turns the 
 {\twist3 $l^\perp_\alpha$} contribution to a  derivative of the \structure{soft term}, of type  
$$
\int d^4z \ e^{- i \ell\cdot z }
\langle \rho(p)|
\psi(0) \,{\twist3 i \, \stackrel{\longleftrightarrow}
{\partial_{\alpha^\perp}}} \bar\psi(z)| 0 \rangle \,.$$
After {\aut Fierz} transformation, this gives the two and three body factorized contributions to the impact factor symbolically shown in Fig.~\ref{Fig:impact-factor-factorization}.
We are thus lead to introduce non-local correlators along the prefered direction $z = \lambda \, \alert{n}$, with  contributions arising from {\aut Taylor} expansion \structure{up to needed term for a given twist order computation}, here 7 correlators at twist 3, which are non-minimal.
These correlators satisfies two equations of motion. Additionaly, the independence with respect to the choice of the vector
  $n$ defining
\bei
	\item the light-cone direction $z$: $z = \lambda \, \alert{n}$

	\item the $\rho_T$ polarization vector: $e_T \cdot \alert{n}=0$

\begin{figure}
\psfrag{pp}{\scriptsize $\!\!k_\perp$}
\psfrag{z}{\scriptsize $k_z$}
\psfrag{u}{\scriptsize $\!\!\!k_0$}
\psfrag{np}{\scriptsize $\!\!\alert{n'}$}
\psfrag{p}{\scriptsize $p$}
\psfrag{n}{\scriptsize $\alert n$}
\qquad 
\centerline{\raisebox{0cm}{\scalebox{.8}{\raisebox{1cm}{$\begin{array}{c}\alert{n}\cdot p=1 \\
\\ n^2=0\end{array}$}\includegraphics[width=2.5cm]{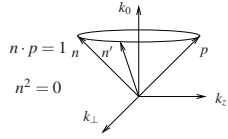}}}}
\caption{Arbitrariness of the light-cone choice for $n$, under the constraint $\alert{n}\cdot p=1$.}
\label{Fig:light-cone-n}
\end{figure}

	\item the axial gauge: $\alert{n} \cdot A=0$
\ei
leads for the amplitude ${\cal A}$ to an equation of the form
\beq
\label{n-ind}
\frac{d{\cal A}}{d\alert{n}_\perp^\mu}
=0
\eq
since only the $\perp$ component of $n$ here matters, as illustrated in Fig.~\ref{Fig:light-cone-n}. 
It can be shown that Eq.~(\ref{n-ind}) implies, for the factorized amplitude ${\cal A}=H \otimes S$, a set of two equations among the non-local correlators. 
Finally, \alert{3 independent DA} are necessary, which are
$\alert{\phi_1(y)}$  (\alert{2-body twist 2 correlator}), 
  $\stmath B(y_1,\, y_2)$ (\alert{3-body genuine twist 3  vector correlator}) and 
$\stmath D(y_1,\, y_2)$ (\alert{3-body genuine twist 3  axial correlator}).

Another approach \cite{Ball:1996tb,Ball:1998sk,Ball:1998ff}, fully covariant but much less convenient when practically computing
coefficient functions, can equivalently be used.  
The dictionnary between these two approaches has been derived 
and explicitly checked for the $\gamma^*_T \to \rho_T$ impact factor at twist 3
\cite{Anikin:2009hk,Anikin:2009bf}. This result, combined to a simple model for the proton impact factor, has been applied successfully  \cite{Anikin:2011sa} to the ratios 
$T_{11}/T_{00}$ and $T_{01}/T_{00}$ measured at HERA.

\subsection{Exclusive $\gamma^{(*)} \gamma^{(*)}$ processes}
\label{SubSec:exclusive-gamma-gamma}

Exclusive $\gamma^{(*)} \gamma^{(*)}$ processes are gold places for testing QCD at large $s$. Aside from studies of the inclusive $\gamma^{*} \gamma^{*}$ total cross-section  \cite{Bartels:1996ke,Brodsky:1996sg,Brodsky:1997sd,Boonekamp:1998ve,Bialas:1997eq,Kwiecinski:1999yx,Kwiecinski:2000zs}, 
there have been indeed several
proposals in order to test perturbative QCD in the large $s$ limit
($t$-structure of the hard $\pom$omeron, saturation, $\odd$dderon...). These are based either on ultraperipheral events where the incoming photon are produced by leptonic or hadronic sources, or on single or double tagged $e^+ e^-$ collisions.
The first proposition was to consider ${\stmath \gamma^{(*)}(q)+\gamma^{(*)}(q^\prime)\to  J/\Psi \, J/\Psi}$, using the mass of the $J/\Psi$ has a hard scale \cite{Kwiecinski:1998sa,Kwiecinski:1999hg}. Then, 
the double tagged lepton scattering at the International Linear Collider (ILC) $e^+ \, e^-\to e^+ \, e^- \rho_L(p_1)+\rho_L(p_2)$ has been proposed and studied
 \cite{Pire:2005ic,Enberg:2005eq,Ivanov:2005gn,Ivanov:2006gt,Segond:2007fj,Caporale:2007vs},
as an access to the subprocess ${\stmath \gamma^*_{L,T}(q)+\gamma^*_{L,T}(q^\prime)\to \rho_L(p_1)+\rho_L(p_2)}$.
These studies have proven the feasibility 
 at {\aut ILC}  of these measurements, based on the expected  high energy and high luminosity of ILC project. A {\aut BFKL} enhancement with respect to {\aut Born} and {\aut DGLAP} contributions is expected\footnote{
Note that this process $\gamma^* \gamma^* \to \rho_L \rho_L$ is dominated at high energy by gluon exchange. At medium energies, quark exchange start to be the dominant contribution, which can be factorized in two ways involving either the GDA of the $\rho$ pair or the $\gamma^* \to \rho$ TDA, depending on the polarization of the incoming photons \cite{Pire:2006ik}.}, with a factor of the order of 4 to 5.

Other proposals have been made, including searches for the elusive $\odd$dderon \cite{Lukaszuk:1973nt}, the $C$-parity odd partner of the $\pom$omeron.
Apart from exclusive tests like $\gamma \gamma \to \eta_c \eta_c$ which only involve the tiny $\odd$dderon exchange \cite{Motyka:1998kb,Braunewell:2004pf}, it has been recently proposed to 
consider the ${\stmath \gamma+\gamma\to \pi^+ \pi^- \pi^+ \pi^-}$ process. Since the
\structure{$\pi^+ \pi^-$ pair has no fixed $C$-parity}, 
$\odd$dderon and $\pom$omeron exchanges can interfere. Thus, although the  $\odd$dderon contribution is presumably tiny, it 
 appears \alert{linearly} in the charge asymmetry \cite{Pire:2008xe}.
However, the distinction with pure QCD processes (with gluons intead of a photon) is tricky, and pile-up at CMS and ATLAS put severe conditions for this measurement.

\section{Conclusion}
\label{Sec:conclusion}

Since a decade, there have been much progress in the understanding of \alert{hard} exclusive processes.   
At medium energies,  \structure{there is now a conceptual framework starting from first principle, allowing to describe a huge number of processes}. 
At high energy, \structure{the impact representation} is a powerful tool for describing exclusive processes in diffractive experiments; they are and will be essential for studying QCD in the hard {\aut Regge} limit ($\pom$omeron, $\odd$dderon, saturation...).
 Still, \structure{some problems remain}.
 	\structure{Proofs of factorization have been obtained only for very few processes} 
(ex.: $\gamma^* \, p \to \gamma \, p\,$, $\gamma^*_L \, p \to \rho_L \, p\,.$)
For some other processes factorization is highly plausible, but not fully demonstrated at any order (ex.: processes involving GDAs and TDAs).
Some processes explicitly show signs of breaking of factorization
(ex.:  $\gamma^*_T p \to \rho_T p$ which has end-point singularities at Leading Order).
Besides, models and results from the lattice for the non-perturbative correlators entering GPDs, DAs, GDAs, TDAs
 are needed, even at a qualitative level!
The effect of \structure{QCD evolution}, the \structure{NLO corrections}, the choice of  \structure{renormalization/factorization scale}, \structure{power corrections} will be very relevant to interpret and describe the forecoming data. 
	At high energy and high luminosity colliders ({\aut LHC, Tevatron, ILC}) exclusive processes are and will be essential for studying QCD in the hard Regge limit ($\pom$omeron, $\odd$dderon, saturation effects...).
To conclude, one should notice that \structure{links between theoretical and experimental communities
involved in exclusive processes are very fruitful}.

\section*{Acknowledgements}

I would like to thank B.~Pire and L.~Szymanowski for discussions and comments.
\vspace{.5cm}

%

%

\bibliographystyle{spphys}       

\end{document}